%% file: OptTech_min.tex
\pdfoutput=1
\documentclass[epjST]{svjour}

\usepackage{graphics}
\usepackage{amsmath}
\usepackage{amsfonts}
\begin{document}
\title{Optical techniques for Rydberg physics in lattice geometries}
\subtitle{A technical guide}
\author{J.B. Naber\inst{1}\fnmsep\thanks{\email{j.b.naber@uva.nl}} \and J. Vos\inst{2} \and R.J. Rengelink\inst{3} \and R.J. Nusselder \inst{1} \and D. Davtyan\inst{1} }
\institute{Van der Waals - Zeeman Institute, Institute of Physics, University of Amsterdam, Science Park 904, 1098XH Amsterdam, Netherlands\and Institute of Quantum Electronics, ETH Z{\"u}rich, 8093 Z{\"u}rich, Switzerland \and LaserLaB, Department of Physics and Astronomy, Vrije Universiteit Amsterdam, De Boelelaan 1081, 1081HV Amsterdam, Netherlands}
\abstract{
We address the technical challenges when performing quantum information experiments with ultracold Rydberg atoms in lattice geometries. We discuss the following key aspects: (i) The coherent manipulation of atomic ground states, (ii) the coherent excitation of Rydberg states, and (iii) spatial addressing of individual lattice sites. We briefly review methods and solutions which have been successfully implemented, and give examples based on our experimental apparatus. This includes an optical phase-locked loop, an intensity and frequency stabilization setup for lasers, and a nematic liquid-crystal spatial light modulator.   
} 
\maketitle
%





\include{intro}	
\include{raman}

\include{locking}						
\include{SLM}

\section{Acknowledgments}
The authors acknowledge support  by the Foundation for Fundamental Research on Matter (FOM), which is part of the Netherlands Organisation for Scientific Research (NWO). JN acknowledges financial support by the  Marie Curie program ITN-Coherence (265031). We especially thank Rick van Bijnen for fruitful discussions, advice and for providing his SLM software as a starting point for our experiments. We also acknowledge Henning Labuhn and the Institue d'Optique Graduate School (France) for their advice on using SLMs with high NA optics. Those collaborations have been greatly facilitated by the Marie Curie Coherence network. We thank Robert Spreeuw and Graham Lochead for discussions and comments on the manuscript, and Jana Pijnenburg, Matthijs Meijers and Tony Hubert for their contributions to the experimental apparatus.

%

\bibliographystyle{ieeetr}
\bibliography{literature} 


\end{document}

%% file: intro.tex






\section{Introduction}\label{ch:intro}
The field of cold atoms has rapidly developed throughout the recent decades. Single atoms can be trapped, cooled and coherently manipulated. Cold atoms qualify as a possible platform for quantum information \cite{Saffman:2010du,Cote:2001uo} and simulation experiments \cite{Weimer:2010vq}, in which qubit information needs to be stored in the atomic states, and a mechanism of entanglement has to be introduced. In this context, exciting atoms to high lying Rydberg states has become increasingly studied \cite{Saffman:2010du}.\par
Any quantum information platform has to provide two basic properties: Initialize and coherently manipulate qubits, and establish entanglement between qubits. The first is usually addressed by preparing the population in selected states of the atomic ground state manifold. Atomic ground states have a long lifetime and are relatively insensitive to environmental influences, such as magnetic or electric fields \cite{Webster:2013tr}. Therefore, qubit information can be coherently stored for up to several seconds \cite{Treutlein:2004ft}. However, when excited to states of high principal quantum number $n$, called Rydberg states, atoms usually experience stronger environmental influence. Many properties scale strongly with $n$, for example, the electric dipole moment ($\propto n^2$), van der Waals interaction strength ($\propto n^{11}$), atomic radius ($\propto n^2$) and lifetime ($\propto n^3$) \cite{Gallagher:1994hu}. As individual addressing of qubits is desirable, the interaction must be strong over distances larger than the size of the addressing laser beam, usually several micrometers. Such interaction strengths between Rydberg atoms can be achieved by using states of high $n$ \cite{Dudin:2012hm} or by the use of F{\"o}rster resonances at lower $n$ to increase the interaction strength \cite{Nipper:2012jv,Vogt:2006kw}. Rydberg states of such high quantum number possess long lifetimes ($>10\,\mu$s) \cite{He:1990dp}, allowing coherent excitation with sufficiently low spontaneous decay rates \cite{Gaëtan:2009fq}. Based on those properties, a mechanism for entanglement was introduced for Rydberg atoms: the dipole blockade. The dipole blockade allows two atoms to be entangled, and to perform gate operations on them \cite{Zhang:2010gk,Isenhower:2010uq}.\par
Several platforms are used to confine cold alkali atoms for subsequent optical manipulation and Rydberg excitation: optical lattices \cite{Schauß:2012ee,Viteau:2011ik}, dipole traps \cite{Urban:2009jd} and magnetic traps \cite{Gunter:2013fv}. All of those implementations pose individual challenges, and have been the focus of intense scrutiny over the recent decades. However, in the context of Rydberg excitation they share common properties and have to tackle common technical problems. Those can be roughly classified in three categories: (i) manipulating the atomic ground state, (ii) exciting the atoms to a Rydberg state, and (iii) the spatial addressing of Rydberg atoms. Transition frequencies in (i) are usually in the microwave or radio-frequency regime, whereas in (ii) transitions are usually optical. The properties of Rydberg states impose properties on the optical transitions. Their long lifetimes imply narrow natural linewidths, typically in the range of several tens of kHz. Therefore, the frequency stability and linewidth of the excitation lasers are of the utmost importance. Furthermore, the intensity of the lasers should ideally be stable over time, as it not only influences the Rabi frequency of the transition, but also shifts the atomic levels via the AC-Stark effect. These shifts lead to an additional dephasing mechanism \cite{Walker:2012bi,Urban:2009jd}. Category (iii) requires single atoms or qubits to be manipulated independently by laser light, as well as addressing multiple atoms at the same time.\par
In this paper we present technical solutions to the challenges arising within the mentioned categories. This involves insights from different fields, such as optics, electronics, control theory and interferometry. The paper is organized as follows: In section 2 we address category (i) by reviewing typical techniques and presenting our realization of an optical phase-locked for two diode lasers. In section 3 we show a scheme for stabilizing and analyzing the frequency and intensity of two diode lasers for Rydberg excitation as required for category (ii). Section 4 gives a brief overview of existing concepts for tackling the problems of category (iii), complemented by a discussion of the technical challenges arising from the use of a nematic liquid-crystal spatial light modulator.

%% file: raman.tex






\section{Ground state atoms and qubits}\label{ch:ground}
\begin{figure}
\includegraphics{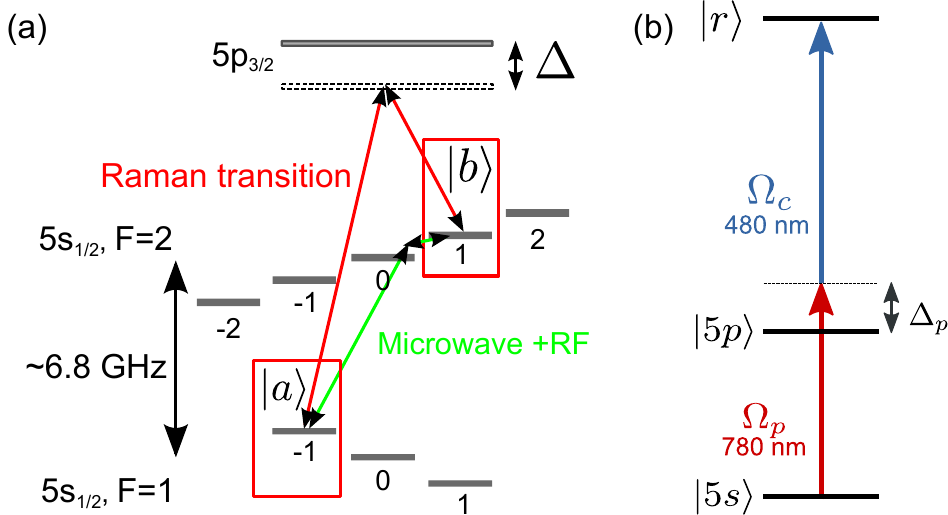}
\centering
\caption{(a) Zeeman sub-levels of the $^{87}$Rb ground state, with the candidate qubit states $|a\rangle$ and $|b\rangle$, which are less sensitive to magnetic field fluctuations at the magnetic magic field. Possible excitation mechanisms between these states use two photons with MW-and RF-radiation, or an optical Raman transition involving two optical photons at $780\,$nm. (b) Two-photon excitation to a Rydberg state $|r\rangle$ using two off-resonant lasers in a ladder type arrangement.}
\label{fig:qubit}
\end{figure}

The hyperfine structure of atoms offers the option to encode qubit information in two different levels of the atomic ground states. These states are long-lived and insensitive to many environmental influences, such as DC-electric fields, thus in principal long coherence times can be achieved. Transitions between the hyperfine-manifolds are magnetic dipole transitions and are usually in the GHz range. Depending on the choice of levels, a qubit transition can be directly driven using a microwave (MW) field \cite{Bize:1999gm} or a two-photon process using MW and radio-frequency (RF) fields \cite{Treutlein:2004ft}. Fig. \ref{fig:qubit} shows the ground state Zeeman manifold of $^{87}$Rb with a spacing of $\sim6.8\,$GHz between the hyperfine levels \cite{Steck:0vg}. The transition between $|F=1,m_F=0\rangle \rightarrow |F=2,m_F=0\rangle$ qualifies as a clock transition, as its frequency is insensitive in first order to magnetic field fluctuations. However, they are magnetically non-trappable states and therefore cannot be used in magnetic traps. Another candidate is the $|F=1,m_F=-1\rangle \rightarrow |F=2,m_F=1\rangle$ transition between magnetically trappable states, which is less sensitive to magnetic field fluctuations around the so called magic field of B=3.23G \cite{Treutlein:2004ft}. As this transition requires a change of $\Delta m_F=2$, it can only be achieved in a two-photon process. We drive this transition using a combination of MW and RF fields, where the individual frequencies are detuned from the Zeeman states (see Fig. \ref{fig:qubit}). The atomic population is prepared in the $|F=1,m_F=-1\rangle$ state using optical pumping \cite{HAPPER:1972ed} prior to the qubit operation.\par
For single-site addressing of individual traps with a few micrometer spacing, one has to refrain from using (MW) radiation alone and implement an optical or optically assisted scheme instead \cite{Weitenberg:2011gn}. A standard solution for optical transitions between alkali atom ground states is to use a Raman transition, where two individual lasers are detuned from an excited state with their frequency difference equal to the ground state splitting (see Fig. \ref{fig:qubit}). Being coupled off-resonantly to the excited state, the atomic population can be transferred coherently between different hyperfine states. As the ground states are long lived states ($\sim\,$1s), the corresponding transition frequency is very narrow, requiring the Raman lasers to be phase coherent over long timescales. Many different schemes have been implemented to achieve a pair of phase coherent lasers. In \cite{Yavuz:2006gj}, the current of a diode laser is modulated, producing sidebands with the desired frequency spacing, after which the carrier frequency is filtered out by the use of a cavity, leaving both sidebands for excitation. In \cite{Thomas:1982ev}, the 0th and 1st order of a high bandwidth acousto-optical modulator (AOM) are used as the excitation pair. Another solution is to produce a sideband at the desired frequency by an electro-optical modulator (EOM) and then selectively injection-lock another diode laser to that sideband \cite{Feng:2014gz}. \par
Besides those implementations a versatile method exists which is based on extending the concept of phase locking from RF applications to the optical regime: the optical phase-locked loop (OPLL) \cite{Enloe:1965it,Leeb:1982et}. This concept, which is based on stabilizing the relative phase between two laser sources, has also been introduced for semiconductor \cite{Steele:1983kq} and grating stabilized lasers \cite{Prevedelli:1995tfbaca}. The starting point is the beat-note signal of two lasers (usually called the ``master"- and ``slave"-laser), which can be used as an error signal for the slave-laser to track the frequency and phase of the master-laser. If a frequency offset between the lasers is required, as for our ground state transitions, one has to down-mix the beat-note signal with the desired frequency \cite{Höckel:2008ft}. This is referred to as a heterodyne OPLL.
\subsection{Raman laser phase lock}\label{ch:raman}
Our experimental apparatus aims at driving the $|F=1,m_F=-1\rangle$ $\rightarrow$ $|F=2,m_F=1\rangle$ transition around the magnetic magic field value. The setup is based on a heterodyne OPLL using a MW generator at $6.8\,$GHz. We present the optical and electronic components together with a discussion of the phase lock performance.
\subsubsection{Phase lock setup}\label{ch:ramanOpt}
\begin{figure}
\includegraphics{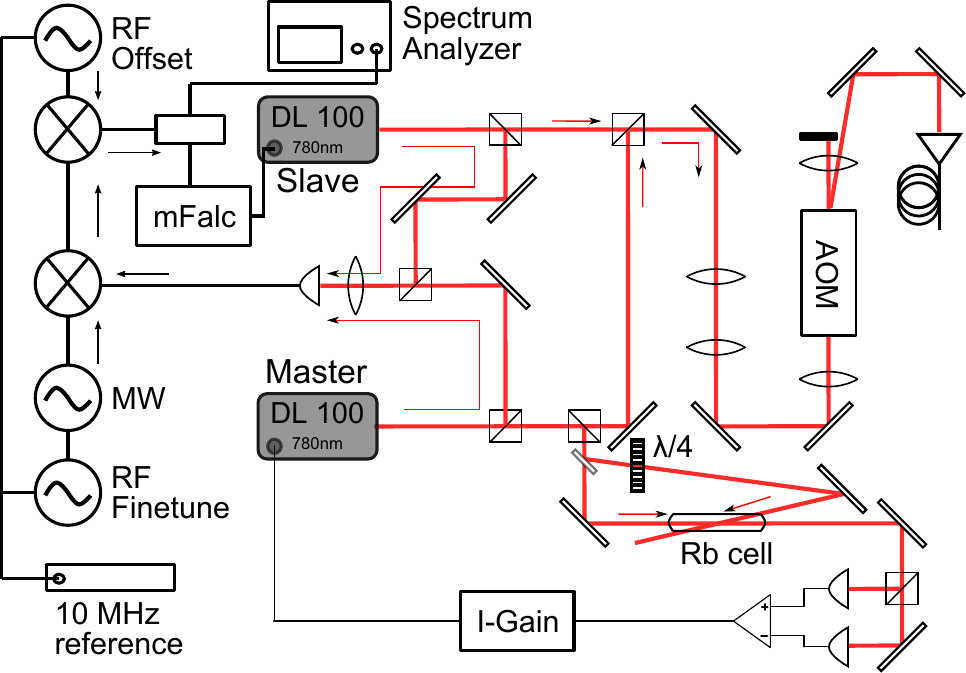}
\centering
\caption{The optical phase lock of the master- and slave-laser at $780\,$nm. The master-laser is frequency stabilized using polarization spectroscopy in a vapor cell. The beat-note signal between the two lasers created at a high-bandwidth photo diode is down-mixed using an analog mixer and a microwave (MW) source. A second mixer creates an error signal using a radio-frequency (RF) generator at $25\,$MHz. The error signal is processed by a commercial PID-controller and fed back to the current modulation input of the slave-laser. Additionally, the error signal is sent via an integral-controller to the piezo element of the laser-cavity for the frequency locking of the slave-laser (not shown).}
\label{fig:raman}
\end{figure}

The optical and electronic setup for phase locking of two commercial diode lasers at $780\,$nm (DL100, Toptica) is shown in Fig. \ref{fig:raman}. The optical components mainly serve to generate a heterodyne beat signal of the two diode lasers around $6.8\,$GHz, where we coarsely adjust the respective laser wavelengths using a commercial wavemeter. The master-laser is frequency locked to a ${}^{85}$Rb transition using the dispersive signal gained from polarization spectroscopy in a rubidium-vapor cell \cite{Pearman:2002kb}. This ensures a stable detuning from the intermediate transition when driving the Raman transition in ${}^{87}$Rb.  We choose polarization spectroscopy instead of frequency-modulation spectroscopy \cite{Borklund:1983vv}, as the necessary frequency sidebands would interfere with the phase lock. Around 20\% of the laser light intensity of each laser is used to generate the beat signal on a high-bandwidth photo diode. The rest of the light is overlapped on a non-polarizing beamsplitter and passes through an AOM before being coupled into an optical fiber going to the experiment. The AOM is used to generate square laser pulses down to approximately 50$\,$ns width.\par
The beat signal is first amplified and then down-mixed at $6.8\,$GHz using an analog mixer. The $6.8\,$GHz signal is generated by a commercial MW-generator, which has been modified such that it is internally phase locked to a commercial, computer-controlled RF-generator. This allows for steering of the MW-frequency with sub-Hz precision. The down-mixed signal is amplified using an RF-amplifier, and split into two parts. One part is compared to a $25\,$MHz frequency reference with a digital phase/frequency discriminator (AD9901, Analog Devices), whose output is low-pass filtered to yield an error signal proportional to the phase difference between signal and reference. We use a low pass filter with a cutoff-frequency of $10\,$kHz, and feed the resulting error signal to the cavity piezo in the slave-laser using an integral-controller. Thereby we achieve a frequency lock of the slave-laser to the master-laser at the desired offset frequency. The other part of the down-mixed signal is again mixed with the $25\,$MHz reference to obtain an error signal, which is processed by a commercial high-bandwidth PID-controller (mFalc, TOPTICA) and fed back to the current modulation input of the slave-laser. This high bandwidth feedback achieves the phase locking.\par
Part of the heterodyne beat signal is split off with a $20\,$dB coupler before the down-mixing and subsequently probed with a spectrum analyzer. This allows for a measurement of the absolute frequency offset between the lasers. Before the second down-mixing with the $25\,$MHz reference, another part of the signal is split off and analyzed by a spectrum analyzer for simultaneous control of the phase lock performance. As all the frequency sources in this setup, including the spectrum analyzer, are phase-locked to a commercial $10\,$MHz rubidium clock, we can get an accurate account of the phase-lock absolute frequency.\par
Any signal delay in the feedback loops leads to a restriction of the achievable feedback bandwidth \cite{Ramos:1990fs}. The two main sources of signal delay are the optical path length of the beat setup and the cable length of the involved electronics. In our setup we try to minimize both lengths, achieving around one meter each. This results in a signal delay of approximately 5$\,$ns.   
\subsubsection{Phase lock performance}\label{ch:phase}
\begin{figure}
\includegraphics{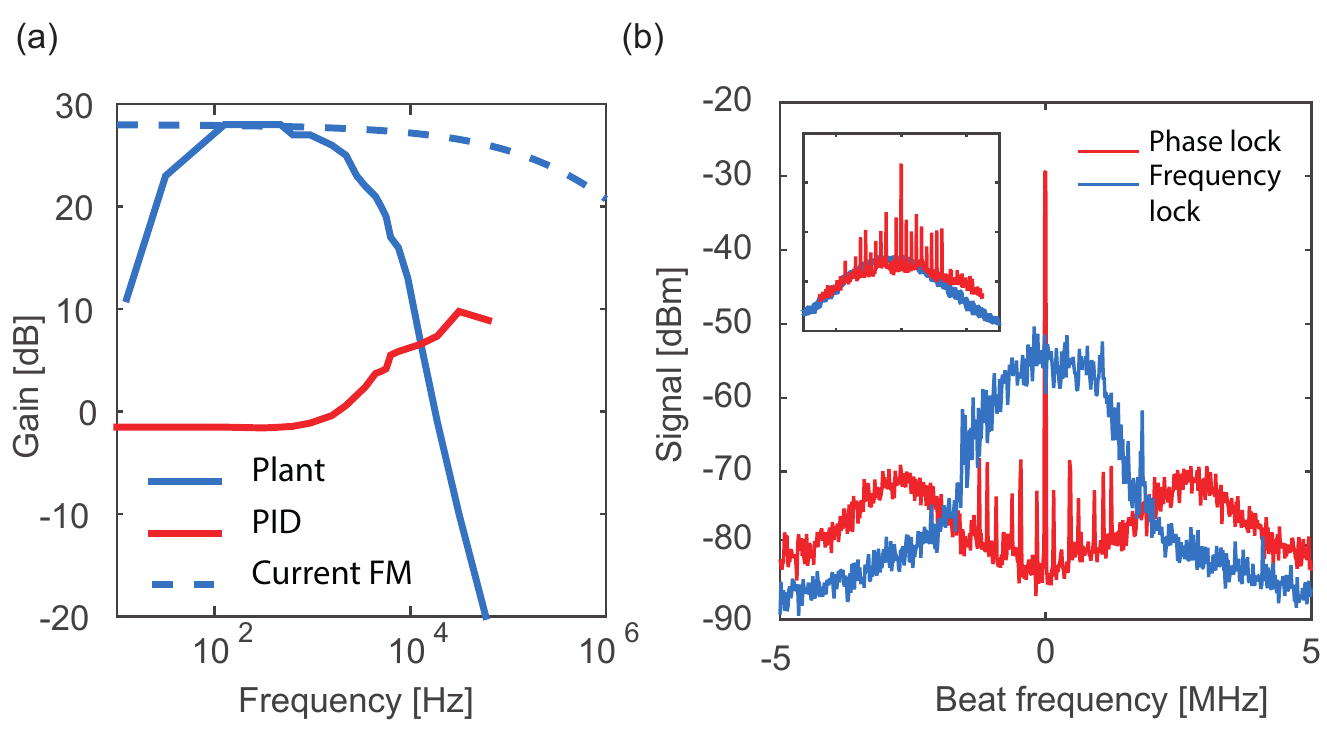}
\centering
\caption{(a) Measured response of the initial loop components to a small sinusoidal input at the laser current modulation input. The bandwidth drops sharply compared to a standard current FM-response (dashed line), hinting at bandwidth limiting components in the feedback loop. Additionally, the PID transfer function is shown for the acquired phase lock. The drop at frequencies $y10^2\,$Hz can be ascribed to the feedback of the frequency lock. (b) Beat signal frequency spectrum referenced to the offset frequency for the final loop design, with a strong improvement in noise suppression as compared to the initial settings (inset).}
\label{fig:raman_note}
\end{figure}

As a rule of thumb, the bandwidth of the feedback system has to exceed the combined linewidth of the lasers, which we find to be around $1.5\,$MHz from the beat signal of the frequency lock in Fig. \ref{fig:raman_note} (b). A fixed limit to the bandwidth is introduced by the PID-Controller (measured bandwidth of $10\,$MHz) and the signal delay. In order to properly evaluate the feedback system's properties, and to subsequently choose proper loop filter settings \cite{Mirabbasi:1999er}, we have to know the transfer function of the involved components. We use the methods which are described in more detail in sections \ref{ch:intens} and \ref{ch:cavity} to evaluate the system's response to a small sinusoidal input on the laser current modulation. To do so, the frequency lock is engaged and the phase-discriminator is used with a higher cutoff-frequency to produce an error signal, which is monitored with a spectrum analyzer. As explained in section \ref{ch:cavity}, when we solely apply feedback to the piezo, the system will only stay part of the time on the linear slope of the error signal. This indirectly dampens the response. To calibrate the overall magnitude we measure the DC response of the system, which can be inferred from the frequency shifts of the laser due to DC current changes and the properties of the error signal. Initially we achieve only a poor phase lock (see inset in Fig. \ref{fig:raman_note} (b)), with insufficient noise suppression and strong spikes resulting from ringing effects in the feedback loop. This can be understood from the frequency response measurement shown in Fig. \ref{fig:raman_note} (a), which reveals a sharp drop at frequencies of approximately $10^4\,$Hz effectively reducing the feedback bandwidth. The reason for achieving phase lock at all is that the PID-Controller partially compensates with increasing gain at high frequencies. 
We use this information to improve the bandwidth of our feedback loop, by removing active components with insufficient bandwidth and significantly reducing the cable lengths and the optical beam path. Introducing additional MW- and RF-amplifiers with low noise figure enhances the open-loop gain and improves the signal-to-noise ratio of the error signal. With those measures the bandwidth is sufficiently increased, such that we achieve a phase-lock with strong noise suppression (Fig. \ref{fig:raman_note} (b)), using mostly the integral and proportional part of the PID-controller to assure very high gain at frequencies of up several MHz. The central feature of the beat spectrum has a linewidth limited by the resolution of the spectrum analyzer ($10\,$Hz), and more than $40\,$dB difference in magnitude to the noise background. To further validate the quality of the phase-lock, we measure the holding and acquisition range. In both cases, we switch off the frequency locking. The holding range is estimated by changing the laser frequency with the piezo element, and observing at which frequency the lock is lost. The acquisition range is measured by setting the slave-laser frequency close to the desired frequency, engaging the feedback and noting the frequency at which the lock is acquired. In both cases we get a range of more than $10\,$MHz, verifying the good loop performance. In general, the frequency lock does not contribute to the phase noise reduction, but it enlarges the holding and acquisition range to several $100\,$MHz, which is indispensable for long term stability.\par
In conclusion, the phase lock provides strong noise suppression and long term stability. Additionally, the offset-frequency can be tuned with sub-Hz precision using the external frequency reference as observed with the spectrum analyzer on the beat signal. The quality of the phase lock proves sufficient to perform transfer of atomic population in our cold atom experiment. Care must be taken in order to ensure sufficient bandwidth in the feedback loop. For an even longer coherence time between the lasers the signal delay must be reduced further, which can be achieved by integrating all electronic components on a PCB and using micro-optics.

%% file: locking.tex






\section{Technical challenges of Rydberg excitation}\label{ch:rydberg}
Ideally, a cold atom or quantum optics experiment delivers monochromatic light with fixed absolute frequency and reproducible intensity. If a system deviates from those properties it will impair the experimental results. For example, in the field of Rydberg physics and quantum information, deviations will lower the fidelity of quantum gates \cite{Zhang:2010gk,Isenhower:2010ke}. The necessary accuracy is governed by the involved atomic levels. Rydberg states of alkali atoms have a lifetime of tens to hundreds of $\mu$s \cite{Gallagher:1994hu,Theodosiou:1984cp}, corresponding to natural linewidths down to $10\,$ kHz. Schemes which excite atoms to Rydberg states in one-, two- or multi-photon processes \cite{Saffman:2010du,Ryabtsev:2011cd} have to provide an absolute and relative frequency accuracy of the order of the natural linewidth of the Rydberg states. Thus, sources of laser light have to be stabilized to that order of magnitude in absolute frequency and linewidth. This precision can usually not be achieved by locking to atomic spectra \cite{Borklund:1983vv,Abel:2009ig}, but necessitates the use of external frequency references.\par
Changes in the laser light intensity can induce changes in the effective Rabi frequency, and thus, for example, lower the contrast in measurements, such as those performed in \cite{Gaëtan:2009fq}. Furthermore, AC-Stark shifts scale with the laser light intensity. Fluctuations in intensity will therefore induce shifts of the atomic levels, which can be influential in dipole traps or during Rydberg excitation \cite{Walker:2012bi}, leading to additional decoherence. Sources of intensity fluctuations are numerous: from the laser source itself, laser beam pointing as well as thermal and acoustic noise in optical fibers. Thus, in some cases active stabilization of the laser light intensity is necessary.
\subsection{Intensity stabilization}\label{ch:intens}
Intensity noise of laser light is introduced by several sources, such as optical fibers, which are subjected to thermal and acoustic noise. This noise also introduces fluctuations in the phase of the laser light, which needs to be compensated in order to deliver laser linewidths down to a few Hz through optical fibers \cite{Ma:1994kx}. In the context of Rydberg excitation this is normally not required as the bandwidth of those noise sources is lower than the natural linewidth of Rydberg atomic levels. Standard methods to stabilize the intensity of laser beams are either passive, such as mechanical rigidity of optical mounts, or active, such as deflecting a variable part of the light by an acousto-optical modulator (AOM) \cite{Burck:2002kv,Takahashi:2008do} or changing the transmittance of an electro-optical modulator (EOM) \cite{Liu:2013ib}. For laser diodes, the current of the diode can be modulated to achieve intensity stabilization. However, this is accompanied by unwanted changes in frequency. \par
\begin{figure}
\includegraphics{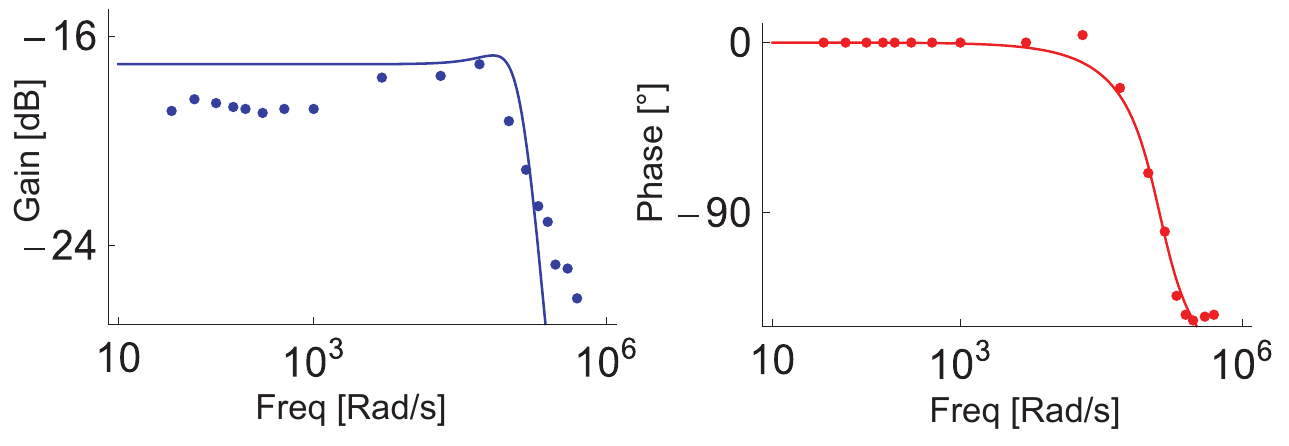}
\centering
\caption{Bode plot of the plant transfer function of the intensity stabilization setup, with measurements (dots) and a theoretical model (line). The measurements are taken from the frequency response of the plant to a small applied sinusoidal signal. The theoretical model is based on a second order system with fitted parameters.}
\label{fig:noise_plant}
\end{figure}

Here we present a setup based on an AOM to stabilize the laser light intensity. We also briefly introduce elements of control theory. They are necessary to understand the principles discussed in sections \ref{ch:phase} and \ref{ch:freqST}, and can easily be evaluated for this system. 
\begin{figure}
\includegraphics{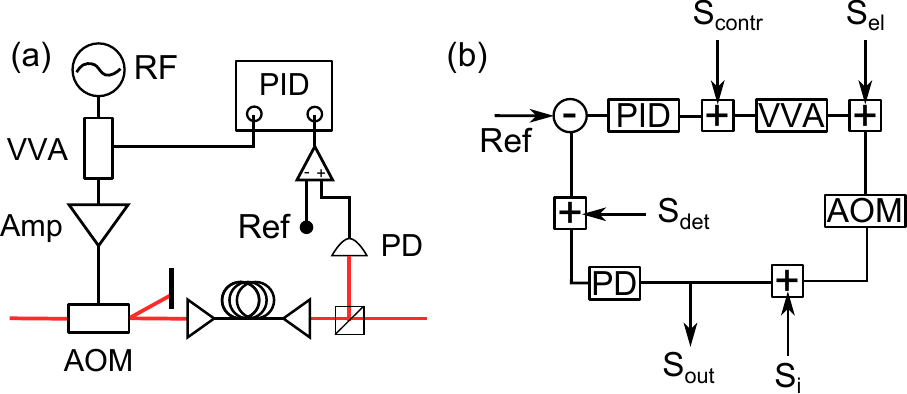}
\centering
\caption{(a) Components of the intensity stabilization setup. The intensity is measured after an optical fiber using a photo diode (PD), guided to a PID-Controller, which controls the RF-amplitude of an acousto-optical modulator (AOM) through a voltage variable attenuator (VVA). (b) Representation of the components in a linear negative feedback loop model. Every block represents the linear transfer function of the respective component. The noise sources present in the system, intensity ($S_{\text{i}}$), controller ($S_{\text{contr}}$), electronic ($S_{\text{el}}$) and detector noise ($S_{\text{det}}$), are added at their respective place in the feedback loop.}
\label{fig:noise_setup}
\end{figure}
Figure \ref{fig:noise_setup} (a) shows the intensity stabilization setup. A variable fraction of the light is deflected into the first order of an AOM and blocked while the zeroth order is coupled into an optical fiber. After the fiber, part of the beam intensity is guided to a photo diode by a beam sampler, and compared to a reference level. The resulting error signal is processed by a PID-Controller and fed back to a voltage-variable-attenuator, which modifies the RF-Power driving the AOM, thereby stabilizing the light intensity in the zeroth order. This setup is an electronic negative feedback system. The characteristic curve of control voltage against intensity is locally sufficiently linear to ensure that linear control theory can be applied. For the description of our system and the subsequent analysis we follow \cite{Ogata:2002tz}. A simple representation of our system in terms of control theory is shown in Fig. \ref{fig:noise_setup} (b). Each element (VVA: voltage-variable attenuator, PD: photo diode, PID: PID-Controller) can be described by an individual transfer function: $X_{\text{VVA}}$, $X_{\text{AOM}}$, $X_{\text{PD}}$ and $X_{\text{PID}}$ with [W${}_{RF}$/V], [W${}_{Light}$/W${}_{RF}$], [V/W${}_{Light}$] and [V/V] as their respective unit conversion (V: Volts, W: Watt). Usually, the system under scrutiny is split into a fixed element, referred to as the plant, and an element whose properties can be changed to optimize the feedback system, here the PID-controller. We can ascribe a transfer function to the plant, $X_{\text{Plant}}=X_{\text{VVA}}\cdot X_{\text{AOM}}\cdot X_{\text{PD}}$, which can be measured. For our system, the open-loop transfer function is given as $X_{\text{OL}}=X_{\text{Plant}}\cdot X_{\text{PID}}$. The system comprises several noise sources, which can be added to the feedback system at their respective position (compare Fig. \ref{fig:noise_setup} (b)) with their noise spectral density. Here we identify electronic noise from the controller ($S_{\text{contr}}$ [V$^2$/Hz]), electronic noise from the RF-source, the VVA and the amplifier ($S_{\text{el}}$ [W$^2$/Hz]), intensity noise from the laser and the fiber ($S_{\text{i}}$ [W$^2$/Hz]) and noise from the photo detector ($S_{\text{det}}$ [W$^2$/Hz]). As we are interested in reducing the intensity noise, we want to know the noise spectral density ($S_{\text{out}}$) of the closed loop. This can be expressed, following the treatment of noise sources in \cite{Ogata:2002tz}, as
	
	\begin{multline*}
	S_{\text{out}}=\frac{1}{(1+X_\text{OL})^2}\\\times\left( S_{I}+S_{\text{el}}\cdot X_\text{AOM}^2+S_{\text{contr}}\cdot X_{\text{VVA}}^2\cdot X_{\text{AOM}}^2+S_{\text{det}}\cdot X_{\text{PID}}^2\cdot X_{\text{VVA}}^2\cdot X_{\text{AOM}}^2\right).
	\end{multline*}
If we assume a high open loop gain, $X_{\text{OL}}\gg 1$, which is necessary to suppress all the additional noise sources, this expression can be approximated as 
\[S_{\text{out}}=S_{\text{det}}\cdot\frac{1}{X_{\text{PD}}^2}.\]
Consequently, with optimal loop performance, the closed loop intensity noise is just limited by the noise performance of the photo detector. We measure the noise in between photo detector and PID-controller while blocking the light before the photo diode. The measured white noise part of the noise spectrum is consistent with the specified photo detector noise. To ensure that both efficient noise reduction and closed loop stability can be achieved by properly adjusting the PID-Controller settings, the plant transfer function has to be known. For our system, the transfer function can easily be retrieved from a frequency response measurement. We apply a small sinusoidal signal to the VVA with varying frequency, and measure amplitude and phase of the signal after the photo detector. From that we can retrieve the Bode plot of the plant transfer function shown in Fig. \ref{fig:noise_plant}.
The measured frequency response can be well described in terms of a second order system (similar to a damped mechanical oscillator) with a transfer function
\[X_\text{Plant}(s)=\frac{k}{1+2\eta \frac{s}{\omega_n}+\left(\frac{s}{\omega_n}\right)^2},\] with fitted parameters $k=0.14$, $\eta=0.6$ and eigenfrequency $\omega_n=1.31\cdot10^5$ rad/s. We can use this theoretical description to find optimal PID-controller settings. Additionally, we need to convert the controller-settings to a transfer function model of type $X_\text{PID}(s)=G_{p}+\frac{1}{T_is}$ (we neglect the differential part in our setup), based on a frequency response measurement from which we extract the proportional gain $G_p$ and the integral time constant $T_i$. Now that we have a theoretical description for $X_\text{plant}$ and $X_\text{PID}$, we can calculate the open loop transfer function (for unity feedback) $X_\text{OL}=X_\text{plant}\cdot X_\text{PID}$ and the closed loop transfer function $X_\text{OL}/(1+X_\text{OL})$. Furthermore, we calculate the system response to a unit step input and retrieve the root locus plot with the proportional gain $G_p$ as a free parameter.\par
\begin{figure}
\includegraphics{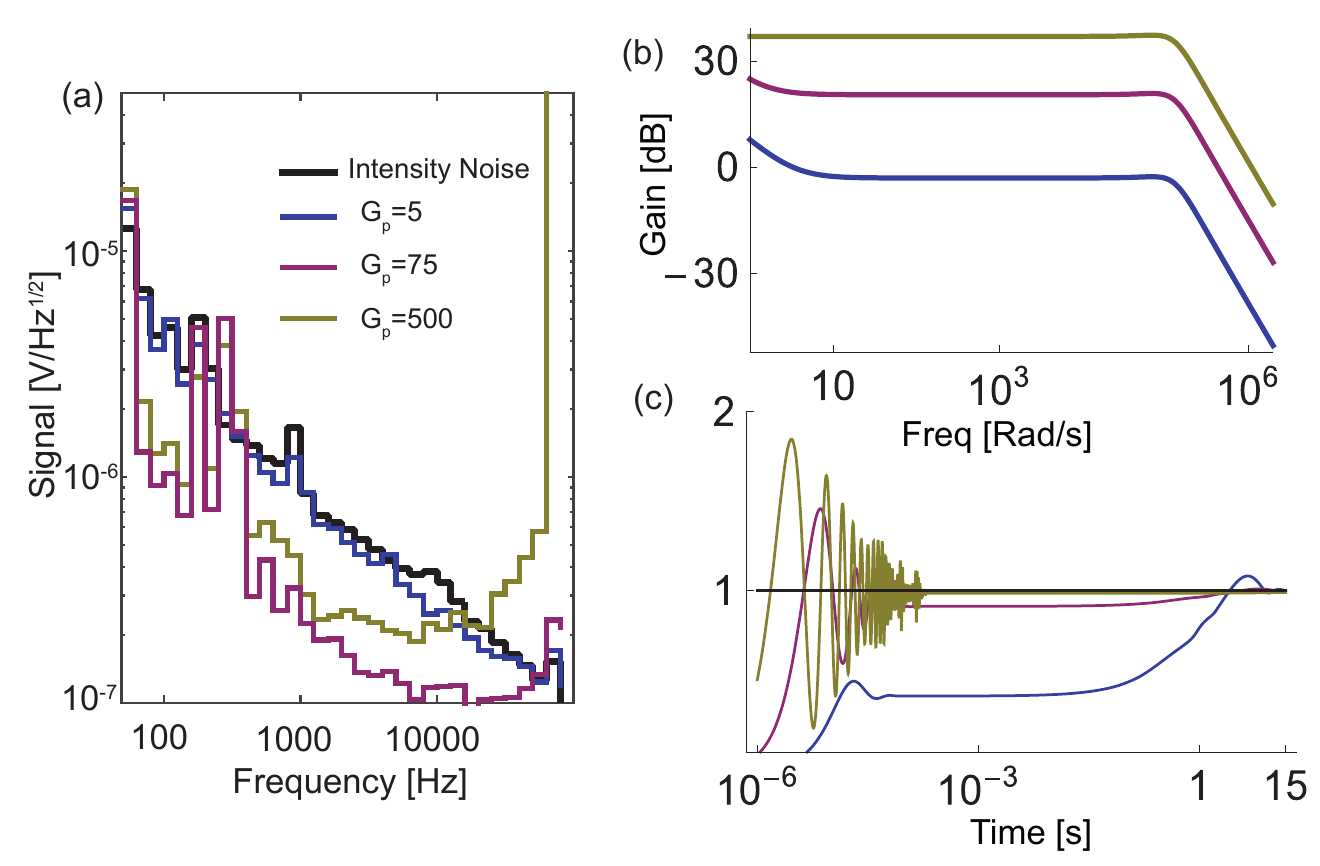}
\centering
\caption{(a) Noise spectral density as measured after the photo diode measured without feedback (intensity noise) and with feedback for different proportional gain values $G_p$. Noise at 50$\,$Hz and odd multiples are related to ground loops between different laboratories. (b) Simulated open-loop gain for the corresponding $G_p$ values. (c) Simulated response of the system to a unit step, showing a highly oscillatory response at high gain, and a slow response for low gain. The time axis was chosen logarithmically to show the different time scales.}
\label{fig:noise_all}
\end{figure}

We measure the noise in the closed loop system after the photo detector using a spectrum analyzer which can measure close to DC signals. From all measurements we subtract the signal which is obtained when the light is blocked. The measured spectrum of the intensity noise, obtained without closing the loop, is shown in Fig. \ref{fig:noise_all} (a).
In the subsequent measurements we close the loop and apply a fixed integral and varying proportional gain. The plot reveals the influence of the feedback system on the intensity noise. For very low proportional gain there is almost no suppression of noise visible. If the gain is increased to 75, there is significant noise reduction over the whole frequency range. Increasing the gain even further leads to a noise increase at high frequencies. The noise at 50$\,$Hz and odd multiples stems from ground loops introduced by differing ground levels between two different laboratories, and can, in principle, be avoided. The feedback system cannot compensate for the uncorrelated ground loop noise. The measurements match nicely to the theoretical evaluation of the loop performance. In Fig. \ref{fig:noise_all} (b) the open-loop gain for the different $G_p$ settings is displayed. For $G_p$=5 the open loop gain is small, insufficient for noise suppression as we demand $X_\text{OL}\gg 1$. The other two gain settings yield sufficient open-loop gain for noise suppression, which is also evident in the measurement. The increased noise at $G_p=500$ can be understood by looking at the system's unit step response in Fig. \ref{fig:noise_all} (c). For high proportional gain the system is highly oscillatory at the eigenfrequency $\omega_n$ of the second-order system, which also shows up as a gain increase in the closed loop at high frequencies. The root locus plot shows that the closed loop poles are moved parallel to the imaginary axis when increasing $G_p$, rendering the system more oscillatory and undamped. Thus, this example illustrates the trade off between sufficient gain for noise suppression and the stability of the feedback system. It should be mentioned that the intensity stabilization mechanism might interfere with laser pulse generation if the bandwidth of both overlaps. In this case, a hybrid solution might be preferable, for example, an AOM for intensity stabilization and an EOM after the optical fiber for pulse generation.

\subsection{Frequency stabilization}\label{ch:freqST}
There are a multitude of different methods to stabilize a laser source to an external reference in the frequency domain: locking to a frequency comb \cite{Inaba:2004hm,Schibli:2005km,Grimmel:2015hr}, using transfer cavity locking \cite{Bohlouli-Zanjani:2006hu}, and, if a stable external laser source is available, beat note locking \cite{Schünemann:1999cv} or injection locking \cite{Mogensen:1985gz}. In many applications optical reference cavities are used \cite{Paldus:2005es,Dahmani:1987it}, as they provide a robust and relatively simple way of reducing the laser linewidth to the Hz \cite{Notcutt:2012ca} and even sub-Hz regime \cite{Salomon:1988fc}. As the locked laser source inherits the frequency stability of that cavity, great care must be taken to limit the frequency drift of the reference. A crucial ingredient is that the material used for mounting the cavity mirrors has a low thermal expansion coefficient \cite{III:1976bp,Takahashi:2012bh}, as any change in cavity length changes the cavity modes' frequencies. Furthermore, the cavity is usually mounted in vacuum using vibration-insensitive components, actively-temperature stabilized and passively temperature shielded. All these measures are meant to reduce environmental influences on the cavity.\par 
As mentioned before, Rydberg excitation requires linewidths of a few kHz, which mitigates the requirements for the reference cavity compared to systems used for optical clocks \cite{Notcutt:2012ca}. Those systems need cavities with high optical finesse (usually $>10^5$), intricate solutions for fluctuations induced by vibrations \cite{Dawkins:2008ef,Keller:2013dm} and are even limited by thermodynamical fluctuations in the mirror coatings \cite{Braginsky:2003cp}.\par
Our external reference is a commercial Fabry-P\'{e}rot cavity (Stable Laser systems) with a finesse of about $2\times10^4$ and free spectral range of 1.5$\,$GHz, which is temperature stabilized around a point of zero thermal expansion. We stabilize two diode lasers onto the cavity using a sideband-locking scheme \cite{Thorpe:2008bv}, achieving linewidths of a few kHz. We measure the absolute frequency to shifts less than 1$\,$MHz within one month. This proves sufficient for the excitation of Rydberg atoms. We present an overview of the optical setup, which is used to stabilize commercial diode lasers, to analyze their linewidths and to get an absolute frequency calibration from spectroscopy. We explain the methods involved and give a brief review about similar approaches.
\subsubsection{Optical setup}\label{ch:RydOpt}
\begin{figure}
\includegraphics{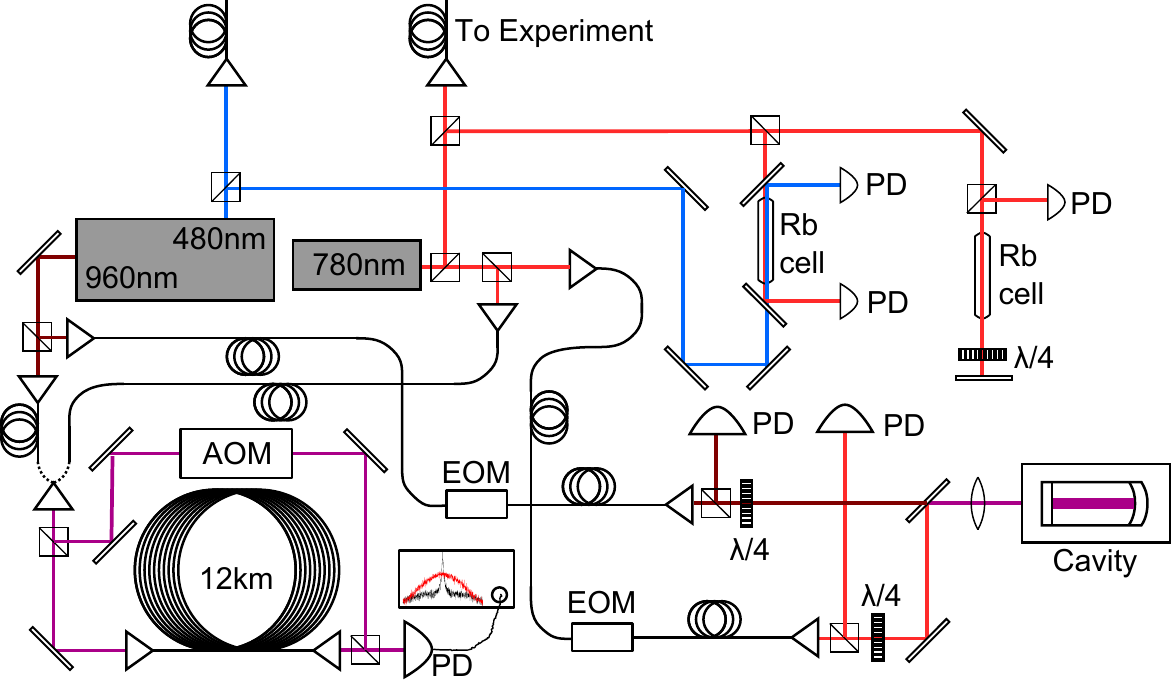}
\centering
\caption{Optical setup for the frequency stabilization and analysis of the 480$\,$nm and 780$\,$nm laser (PD: photo diode, AOM: acousto-optical modulator, EOM: electro-optical modulator, violet beam: shared beam path). The lasers are locked to the same high-finesse cavity using the Pound-Drever-Hall method. The linewidth of both lasers can be measured using a self-heterodyne fiber beat involving a 12$\,$km long telecommunication fiber. The laser light is used for Doppler-free absorption  and EIT spectroscopy on room temperature vapor cells to calibrate the absolute laser frequency.}
\label{fig:locking}
\end{figure}
The optical setup is sketched in Fig. \ref{fig:locking}, consisting of two commercial diode lasers at 780$\,$nm and 960$\,$nm wavelength (DL Pro and TA-SHG Pro, Toptica), the latter of which is frequency doubled to 480$\,$nm. The output powers are 60$\,$mW and 450$\,$mW respectively and the free-running linewidth is roughly 200$\,$kHz. These two lasers are used for the two-photon excitation scheme of $^{87}$Rb as depicted in Fig. \ref{fig:qubit} (b). For both lasers part of the light is guided into a fiber-coupled EOM with high bandwidth (PPM785 and PPM960, Jenoptik), which introduces sidebands through phase modulation of the incoming light. After traveling the fiber the light is out-coupled into the high-finesse cavity passing through mode-coupling optics (a combination of telescopes and an imaging lens). The Gaussian beam profile of the light after the fiber helps to efficiently couple the light to the Gaussian TEM${}_{00}$ mode of the cavity. For the design of our cavity, consisting of a flat and convex mirror, the best mode matching is achieved when the minimal waist of the Gaussian beam matches the waist of the cavity mode at the entrance mirror \cite{Siegman:1986uk}. With this arrangement our coupling efficiency is around 80\%, as measured by the drop of the reflected signal when scanning across the cavity resonance. A combination of dichroic optics and optical bandpass filters ensures that the different wavelengths can be coupled simultaneously into the cavity, whose mirrors are coated for these two wavelengths. Light reflected from the cavity is detected by a fast photo diode, whereas transmitted light is guided to a slow photo diode and a camera for spatial mode investigation. The bandpass filter prevents light from one wavelength leaking into the photo diode dedicated to the other, which would be detrimental for the subsequent error signal generation.\par
Another part of the light at these two wavelengths is fiber coupled to the self-heterodyne fiber beat setup for analyzing the linewidths (see section \ref{ch:linewidth}), which is built with components supporting both wavelengths. Independent examination of the two lasers' linewidth can be achieved by exchanging the input fibers. The light is split into two parts, the first of which is guided to an AOM in double-pass configuration, which was chosen over a single-pass configuration to make the setup independent of the wavelength. The AOM is required in order to shift the laser beat signal away from DC. The other part is coupled into a 12$\,$km long telecommunication fiber, which strongly attenuates the light intensity at the near-infrared wavelengths, thus requiring high input powers of tens of mW to achieve $\mu$W output powers. The out-coupled light is overlapped with the light from the AOM on a photo detector, yielding a beat signal.\par
The rest of the light is unaffected by the modifications made in these two setups (e.g. the sidebands introduced in the EOMs) and is used for Rydberg excitation in the main experiment. A small fraction of the light is branched off in order to simultaneously monitor the laser frequency on a commercial wavemeter, as well as to perform Doppler free absorption spectroscopy at 780$\,$nm in a rubidium vapor cell. This allows us to have an in-situ analysis of the wavelength while performing experiments. Additionally, the light can be switched to enter another rubidium vapor cell in a counter-propagating arrangement, allowing for EIT measurements as described in section \ref{ch:frequency}.
\subsubsection{Cavity locking}\label{ch:cavity}
We stabilize our laser frequency to the high-finesse cavity using the Pound-Drever-Hall (PDH) locking  scheme \cite{Drever:1983do,Black:2001fe}, which uses the response of the cavity in reflection \cite{Houssin:1990jr} as recorded by the photo diodes in Fig. \ref{fig:locking}. To obtain the typical error signal of the PDH, sidebands at 2$\,$MHz are introduced onto the light by means of the fiber-coupled EOM. The signal of the photo diodes is demodulated with the sideband frequency using an analog mixer, yielding the error signal. The modulation depth of the sidebands is tuned to increase the error signal slope at the zero crossing. The sideband frequency is chosen to allow for broad non-zero frequency ranges around the slope, which assist in acquiring the lock. \par
If we stabilize our laser to the cavity modes, which are fixed in frequency space (besides thermal drifts etc.), we naturally encounter the problem that those frequencies do most likely not match the atomic transition frequencies (the free spectral range (FSR) of the cavity is 1.5$\,$GHz). Several solutions to this problem exist. One method is to change the cavity modes' frequency by changing the cavity mirror spacing using piezo-electric transducers. This solution usually entails reduced mechanical rigidity and increased sensitivity to thermal fluctuations. Another approach is to use high bandwidth AOMs or cascades of AOMs to shift the laser frequency, introducing the problem of beam-pointing and reduced laser intensity. Therefore, we choose to use the offset sideband-locking scheme as suggested in \cite{Thorpe:2008bv}, which offers free tunability of the laser frequency and similar noise characteristics as the PDH locking scheme. The basic principle is that instead of locking the carrier frequency of the laser, we use sidebands introduced via the phase modulation of the EOM. By changing the frequency of those sidebands we can, in principle, address any desired frequency, if the bandwidth is sufficient to cover at least half the FSR of the cavity. In practice a higher bandwidth is beneficial. For ideal phase modulation, we optimize the modulation depth to get maximum power in the first order sidebands, which we use for locking. In order to achieve sideband locking with the PDH mechanisms we combine the frequency of the locking sidebands and the frequency of the PDH sidebands using a combiner, and send both to the EOM. This allows for creating an error signal at the offset sideband frequency.\par
\begin{figure}
\includegraphics{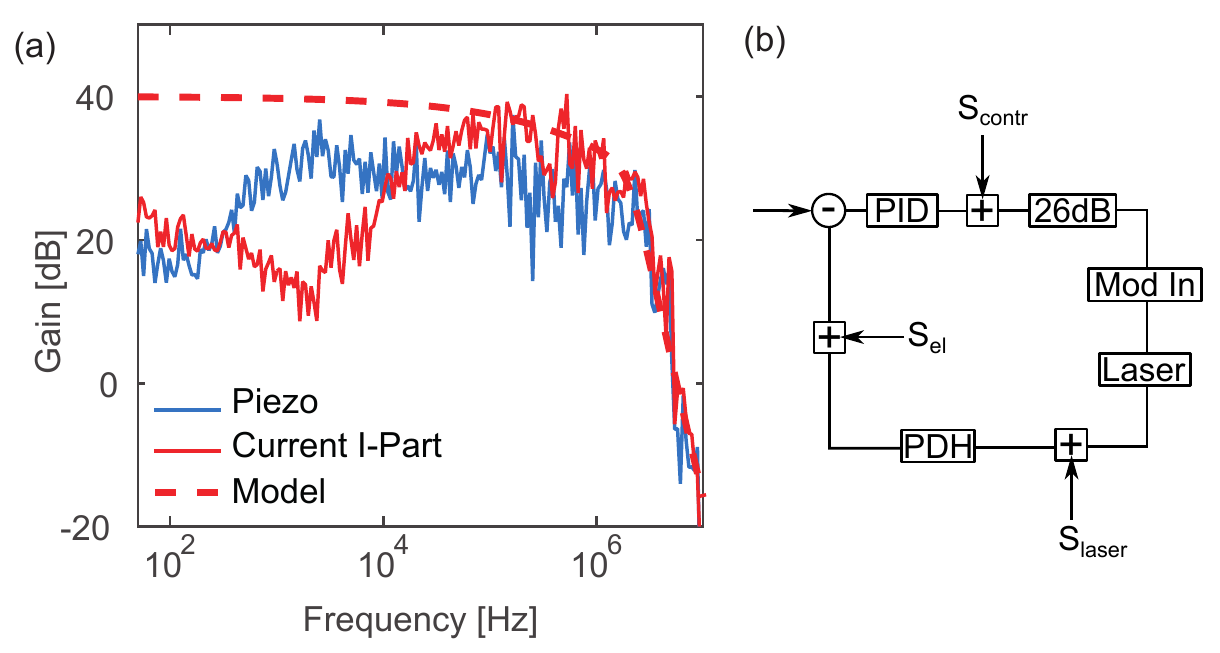}
\centering
\caption{Gain measurement of the feedback loop transfer function using either feedback to the laser cavity piezo or solely involving the integral part of the laser current feedback. This leads to a partial lock of the laser, influencing the transfer function up to the respective bandwidth of piezo and integral part ($10^3\,$Hz and $10^5\,$Hz). The dashed lines shows the current FM-model described in the text. (b) Representation of the feedback loop components (26$\,$dB: attenuator, Mod In: laser current modulation input, PDH: all components involved in the PDH error generation. The main noise sources are laser phase noise ($S_\text{Laser}$), electronic noise ($S_\text{el}$) and controller noise ($S_\text{contr}$).}
\label{fig:transfer780}
\end{figure}
Similar to the analysis in section \ref{ch:intens}, we need to investigate the feedback system for optimal stabilization and noise suppression. The main source of noise is the laser linewidth, which is translated into 
voltage noise by the PDH mechanism. Fig. \ref{fig:transfer780} (b) shows a simplified representation of our feedback systems. For simplifying the representation, the PDH module in the representation comprises several independent components like the cavity, photo diodes and the electronics of the demodulation. All of those components have an individual transfer function, which we do not analyze independently for simplicity. The error signal is compared to 0$\,$V for a perfectly symmetric error signal and the difference signal is fed to a commercial digital PID-controller (Digilock, Toptica), where the error signal is split into two independent PID-modules. The integral component of one module is used to control the voltage of the laser piezo element to compensate for long term drifts in laser frequency. The bandwidth here is limited to a few 100$\,$Hz. The other module is used to control the laser frequency through the modulation input of the laser after 26$\,$dB of attenuation, which converts a change in voltage into a change of laser current and thereby frequency. Due to the higher bandwidth of several MHz, this part is responsible for the linewidth reduction of the laser. At the input of the current module, we engage a third-order bandpass filter with a 2$\,$MHz cutoff-frequency. All those elements are described by the transfer functions  $X_{\text{PID}}$, $X_{\text{26\,dB}}$, $X_{\text{Mod In}}$, $X_{\text{Laser}}$ and $X_{\text{PDH}}$ with [V/V], [V/V], [A/V], [Hz/A] and [V/Hz] as their respective unit conversion. The dominant noise sources are the laser linewidth, entering as $S_\text{Laser}$, the electronic noise $S_\text{el}$ and controller noise $S_\text{contr}$.\par
As shown in section \ref{ch:intens}, a high open loop gain suppresses the noise in the closed loop system, and thereby reduces the laser linewidth. To achieve a high open loop gain for the whole bandwidth of laser noise, we need to know the plant transfer function $X_\text{plant}=X_{\text{26\,dB}}\cdot X_{\text{Mod In}}\cdot X_{\text{Laser}}\cdot X_{\text{PDH}}$ to adjust the PID parameters accordingly. However, a frequency response measurement similar to the one for the intensity stabilization is not straightforward to implement, as the system does not reside in the linear part of the PDH error signal without applying feedback. In principle, the response of each independent component can be measured, for example the frequency modulation (FM) response of the laser to the laser current, or the characteristics of the PDH module. This adds additional complexity, so we choose for a simpler approach instead. We engage either the piezo-feedback or the I-Part of the current feedback, which results in the system oscillating back and forth around the slope of the error signal, as the feedback is too weak to ensure tight locking. Then we introduce a small sinusoidal signal, add it to the error signal and observe the amplitude of the response with the built-in network analyzer of the PID-controller. The measurement for the 780$\,$nm-laser is shown in Fig. \ref{fig:transfer780} (a), which clearly features the bandwidth of the relative feedback mechanisms as suppression of the signal at low frequencies. We compare it with a theoretical prediction combining the FM-response of the laser described by the absolute value of
\[X_\text{FM}(i\omega)=-\frac{C}{3}\times\frac{3-\sqrt{i\frac{\omega}{\omega_c}}}{1+\sqrt{i\frac{\omega}{\omega_c}}}\] as an empirical model for the carrier-induced and thermal FM \cite{Kobayashi:1982ij,Saito:1984gp,Corrc:1994bc} (C and $\omega_c$ are fit parameters, the parameter b in \cite{Corrc:1994bc} was chosen to be 3), and the transfer function of the third order bandpass filter at 2$\,$MHz. The model accurately describes the system response in the range where the signal is not affected by the relative feedback mechanism. Furthermore, we retrieve the DC-response of our system by calibrating the laser frequency shift at DC-inputs to the current modulation and by measuring the slope of the error signal. The measured DC-gain of around 40$\,$dB is consistent with the model prediction. The high gain of the plant transfer function explains why the laser can be locked by just feeding back the error signal to the laser current, without involving any PID-controller.\par
\begin{figure}
\includegraphics{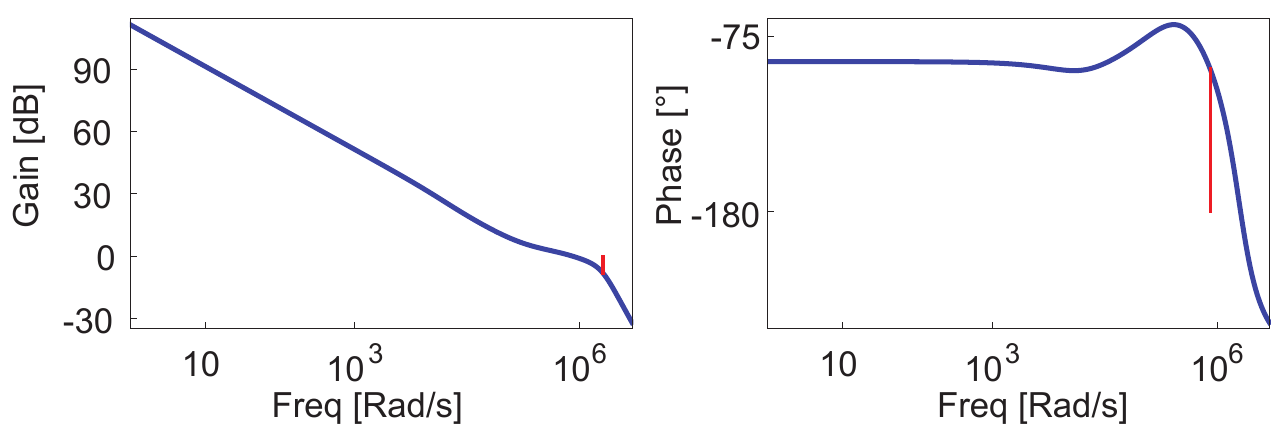}
\centering
\caption{Bode plot of the frequency stabilization open-loop transfer function. The Bode plot is calculated from the measured transfer function of the plant and the PID-controller. Red bars indicate the positive gain- and phase-margins respectively, which are a criterion for the closed-loop stability.}
\label{fig:OL780}
\end{figure}
Additionally, we calibrate the PID-module settings with a frequency response measurement, converting it to the standard form $X_\text{PID}(s)=G_p+\frac{1}{T_is}+T_ds$. Now we can calculate the open-loop transfer function $X_\text{OL}=X_\text{PID}\cdot X_\text{plant}$ for our locking parameters. The optimal set of parameters can be retrieved empirically from the measurements in section \ref{ch:linewidth}, yielding a laser linewidth of less than 10$\,$kHz. For this set of parameters the open-loop transfer function with the respective phase and gain margins is plotted in Fig. \ref{fig:OL780}. From the plot, we can understand the loop's ability to efficiently suppress noise. A Fourier transform analysis of the error signal shows that the laser frequency noise is limited in bandwidth to 500$\,$kHz, with dominant contributions at low frequencies. High open loop gain at those low frequencies is necessary to suppress the noise and is provided by the I-Part of the feedback, which shows up as the -20$\,$dB linear slope per decade at low frequencies. The third order bandpass filter at 2$\,$MHz allows for sufficient noise suppression at low frequencies, and filters uncorrelated high frequency noise. The steep slope of the filter provides sufficient phase margin despite high overall gain. If we increase the gain even further, the small gain margin vanishes completely, and the system becomes oscillatory, which is clearly evident on the error signal and the heterodyne-fiber beat measurement.
\subsubsection{Absolute frequency calibration}\label{ch:frequency}
An important aspect of optical experiments is to get an absolute frequency reference for spectroscopic experiments. Several methods exist, for example, referring the laser frequency to a frequency comb \cite{Takamoto:2005ef}, or using a commercial wavemeter. Commercial wavemeters offer a relative accuracy up to a few tens of MHz, which usually requires frequent recalibration or the use of built-in reference lasers. Even if the laser frequency is known with sufficient accuracy, the theoretical prediction of the atomic transition frequency might not match that accuracy. A simple way to find atomic transition frequencies for cold atom experiments is to use spectroscopy of room temperature vapor cells, as shown in Fig. \ref{fig:locking}. First, we perform Doppler free saturation spectroscopy with the 780$\,$nm laser, to find the 5s$_{1/2}$ $\rightarrow$ 5p$_{3/2}$ transition  frequencies of $^{87}$Rb. Then, we fix a sideband of the offset-locking scheme such that the laser frequency matches the transition frequency. We are left with finding the 480$\,$nm-laser frequencies corresponding to 5p$_{3/2}$ $\rightarrow$ ns$_{1/2}$ and nd$_{3/2,5/2}$. A common scheme is to use electromagnetically-induced transparency (EIT) \cite{Fleischhauer:2005ix} with two on-resonant lasers in a ladder-type excitation scheme. EIT spectroscopy can yield sub-MHz precision of Rydberg transition frequencies \cite{Mack:2011he}, and can also be used to stabilize the 480$\,$nm laser \cite{Abel:2009ig}.\par
\begin{figure}
\includegraphics{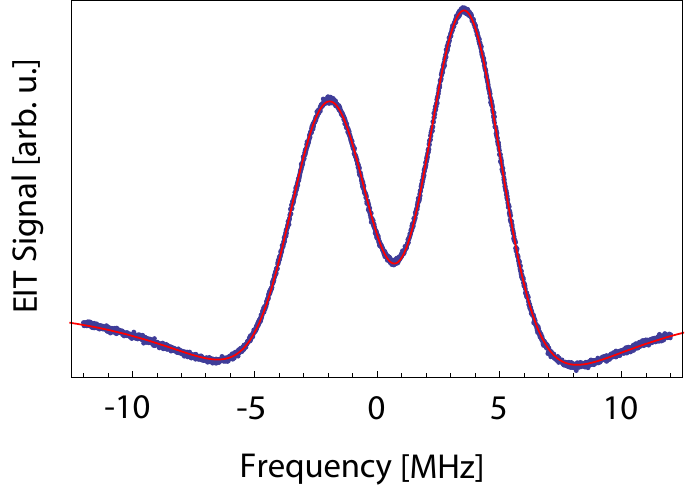}
\centering
\caption{Electromagnetically induced transparency (EIT) as measured by the transmitted 780$\,$nm light while scanning the frequency of the 480$\,$nm laser. The red line shows a fit to the data using an analytical model as mentioned in the text. The measurements shows the two hyperfine levels (F=1,2) of the 22s$_{1/2}$ state with sub-MHz resolution.}
\label{fig:EIT}
\end{figure}

Our setup is similar to the one described in \cite{Tauschinsky:2013iq}, featuring two counter-propagating laser beams in a vapor-cell at 480$\,$nm and 780$\,$nm. The vapor-cell is magnetically shielded using mu-metal to minimize systematic deviations due to Zeeman splitting of atomic lines. The transmitted 780$\,$nm laser light is recorded with a photo diode, with maximal absorption if there is no 480$\,$nm light present. If the frequency of the 480$\,$nm laser is scanned such that it matches the Rydberg transition, the medium is rendered partially transparent, increasing the transmitted 780$\,$nm light intensity. We scan the laser frequency by sweeping the sidebands of the offset-locking scheme. If the slew rate of that sweep is sufficiently smaller than the bandwidth of the locking system, the locked laser follows the sideband frequency smoothly. The signal-to-noise ratio is largely improved by chopping the 480$\,$nm light and using lock-in detection, yielding Rydberg spectra as shown for the 22s$_{1/2}$ state in Fig. \ref{fig:EIT}, with sufficient resolution to distinguish between different Rydberg hyperfine-states.
We fit our EIT spectrum with the analytical expression found in \cite{Gea-Banacloche:1995ix} (see red line in Fig. \ref{fig:EIT}), with a fitting error usually a few tens of kHz for the transition frequency.\par
In conclusion, EIT measurements in a room-temperature vapor cell allow for sub-MHz accuracy in measuring the transition frequency of Rydberg states. If the on-resonance frequency of both lasers is known from such a measurement, a well defined detuning can be introduced via the offset-locking scheme allowing for two-photon off-resonant excitation as shown in Fig. \ref{fig:qubit} (b).

\subsubsection{Linewidth determination}\label{ch:linewidth}
After introducing the cavity locking scheme we need to verify that the achieved reduction in linewidth is sufficient for the particular application, in our case off-resonant Rydberg excitation, requiring linewidths of around 10$\,$kHz. In principle, the laser linewidth can be deduced from the spectral distribution of the closed-loop error signal \cite{Elliott:1982jm}. We refrain from that approach as we find electronic noise uncorrelated to the laser linewidth in our error signal, most likely originating from leakage of frequencies of the offset-locking into the demodulation of the PDH setup. If two identical laser sources (e.g. two lasers locked to the same reference cavity) or an independent narrow-linewidth reference laser is present, the linewidth can be deduced from a heterodyne beat measurement of those laser sources \cite{Saito:1981fz}. Furthermore, the laser linewidth can be inferred from a time-domain measurement of the Allan deviation of the frequency \cite{Barnes:1971ik}. As we do not have two identical laser sources, and also want to optimize our lock-settings in-situ, we chose a self-heterodyne measurement (see Fig. \ref{fig:locking} and section \ref{ch:RydOpt} for the optical setup), as first introduced in \cite{Okoshi:1980kt}. 

\begin{figure}
\includegraphics{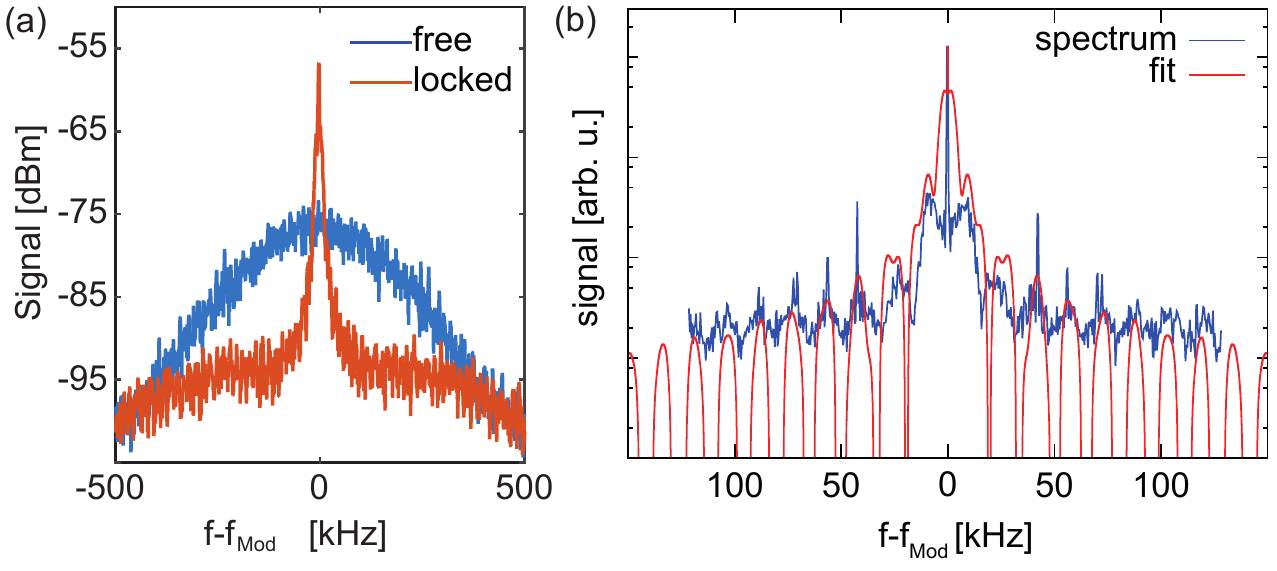}
\centering
\caption{(a) Frequency spectrum of the self-heterodyne beat signal for the free-running 780$\,$nm laser (linewidth 230$\,$kHz) and the same laser locked to the high finesse cavity. (b) High resolution frequency spectrum for the best performing feedback loop of the 960$\,$nm laser. The strong interference fringes indicate that the laser coherence time exceeds the delay time in the 12$\,$km fiber. The spectrum can be fitted using an analytical model based on the assumption of white phase noise, yielding a laser linewidth of 3.2$\,$kHz and a delay time of $65\,\mu$s.}
\label{fig:hetFib}
\end{figure}

As in the case of a heterodyne beat measurement the linewidth is inferred from a beat signal. The delayed light traveling through a long fiber acts as the second laser source. To visualize that idea, one can regard the delayed light as an independent, uncorrelated laser source if the delay time of the light $\tau_0$ in the fiber exceeds the coherence time $\tau_\text{coh}$ of the laser. In this regime ($\tau_0>\tau_\text{coh}$) the resulting beat linewidth can be converted into the laser linewidth by correcting with a factor of 1/2 or $1/\sqrt{2}$ for the case of Lorentzian or Gaussian linewidth. An example for a beat signal in that regime can be found in Fig. \ref{fig:hetFib} (a), for the case of the free running laser at 780$\,$nm and the same laser with feedback. The resulting free running linewidth is measured to be around 230$\,$kHz for a Lorentzian lineshape, consistent with the specifications of that laser. The feedback system yields a significant reduction in laser linewidth, clearly evident in the beat signal. If we optimize our locking parameters such that the resulting laser linewidth drops beneath the inverse delay time $1/\tau_0\approx16$\,$\text{kHz}$, the simple approximation for the laser linewidth obtained from the beat signal breaks down. This case is shown in Fig. \ref{fig:hetFib} (b) for the 960$\,$nm laser, where the spectrum features equidistantly spaced fringes, that become more pronounced if $\tau_\text{coh}\gg\tau_0$. With the assumption that only white noise is present in the phase fluctuations of the laser, yielding Lorentzian line shapes, one can find an analytical expression for the spectral noise distribution in the fiber beat signal \cite{Horak:2006ja}:
\begin{equation*}
\begin{split}
S(f) \propto \frac{1}{\pi} \frac{\Delta f}{(\Delta f) ^2+(f -f_0)^2} \times \\
	     &\left( 1 - e^{-2 \pi \Delta f \tau_0} \left[ \cos(2 \pi \tau_0 (f-f_0)) + \frac{\Delta f \sin(2 \pi \tau_0 (f-f_0))}{f-f_0}\right] \right),
\end{split}
\end{equation*}
where $f_0$ is the modulating frequency of the AOM and $\Delta f$ is the laser linewidth. 
It should be noted that in reality there are additional noise sources on the laser, such as $1/f$ noise at low frequencies. In such cases the relation between noise and linewidth is more complicated \cite{Domenico:2010icbaca}, but in principle, assuming a certain noise distribution, the linewidth can also be inferred from the self-heterodyne measurement \cite{Ishida:1991kk}. However, there is acoustic and electronic noise in the self-heterodyne beat setup itself, especially acoustic noise in the long fiber. Thus, measuring linewidths less than 1$\,$kHz is not feasible in this setup. \par
We restrict ourselves to the analytical expression in fitting the data. We use $\tau_0$, $\Delta f$ and an overall offset as free parameters, and exclude frequencies within -30 to 30$\,$kHz to retrieve acceptable results. The fit yields a realistic fiber delay time of $\tau_0=65 \,\mu$s, a linewidth of $\Delta f=2.3\,$kHz and reproduces the features of the spectrum. From the reasoning above it is clear that this estimate of the linewidth has limited accuracy. However, if we also take into account that the spectral fringes are strongly pronounced, as predicted by theory for $\Delta f\ll 16\,$kHz, we can safely assume to have a linewidth of less than 10$\,$kHz, matching the requirements.

In conclusion, we achieve absolute frequency stabilization and linewidth reduction of two diode lasers, sufficient for off-resonant Rydberg excitation as observed in our cold atom experiment. The decrease in linewidth is also visible in EIT measurements, as shown in Fig. \ref{fig:EIT} compared to similar measurements with free running lasers.

%% file: SLM.tex






\section{Spatial addressing}\label{ch:spatial}
We refer to spatial addressing as the ability of shaping the spatial intensity distribution of laser light, for the purpose of creating trapping geometries or to selectively excite atoms. There exists a multitude of different realizations, either static or dynamic: dipole trap arrays generated by microlens arrays \cite{Dumke:2002hn}, beam steering with acousto-optical modulators \cite{Yavuz:2006gj}, coupling atoms to optical waveguides \cite{Zoubi:2014hz}, using digital mirror devices (DMDs) in a binary amplitude modulation \cite{Akbulut:2011fn,Muldoon:2012jt} or holographic mode \cite{Lee:1974ed,Conkey:2012cr}, projecting a binary mask \cite{Bakr:2009bx}, or using liquid-crystal spatial light modulators (SLM) \cite{Bijnen:2015da,Bergamini:2004bl}.\par
Spatial light modulators in general describe devices which modulate amplitude, phase or polarization of light, based on micro-electromechanical systems (MEMS), diffractive optical elements (DOE) or liquid crystal technology \cite{Lazarev:2012ku,Gunther:2004ix}. DMDs, belonging to the MEMS, are a versatile tool for light modulation, also in commercial applications \cite{Ramanath:2015ua,Knipe:1996gz}. They offer high update rates (in the kHz range) and high optical resolution, up to $10^6$ pixels. Liquid crystal type spatial light modulators have lower update rates of up to 60$\,$Hz, but similar resolution, a high dynamical range and higher efficiency. Our experiment aims at spatial addressing of atoms in a magnetic lattice, both for Raman type transitions and Rydberg excitation. This requires to change the spatial excitation pattern once per experimental cycle ($\sim$ 20$\,$s). Therefore we do not depend on high update rates. Rydberg excitation however depends on high laser power at 480$\,$nm in the two-photon excitation scheme, thus high efficiency of the light shaping is preferred. As a consequence we opt for a SLM based solution. In general, if conversion efficiency is more important than fast update rates, like for dipole trap arrays \cite{Nogrette:2014fj} or Rydberg excitation \cite{Bijnen:2015da}, SLM devices have become increasingly popular in the field of cold atoms for their versatility and ease of operation.\par
Due to intensive research for scientific and commercial applications \cite{Lazarev:2012ku,Serati:2005dg}, SLMs and their working principle are well understood. Usually, the beam shaping is done by means of geometric beam shaping \cite{Bryngdahl:1975jv} or the iterative Fourier transform algorithm  (IFTA) \cite{Wyrowski:1988dh}. Geometric beam shaping is based on dividing the SLM into independent areas of locally different phase gradients for beam steering. This method is simple to implement, but suffers from low efficiency due to the fact that just part of the SLM contributes to a given point in the image plane. The IFTA method utilizes the fact that for light originating from the SLM and being imaged by a lens, the light fields at the SLM and focal plane of the lens are related by a Fourier transform in the Fresnel approximation. For a review of these two methods and an efficient implementation of the IFTA method, see \cite{Bijnen:2013ve}.\par
For the excitation or confinement of atoms in lattice geometries the SLM can be used to create diffraction limited spot patterns. In conjunction with high numerical aperture (NA) lenses, great care has to be taken to produce reproducible spot patterns of homogeneous intensity. As recently shown in \cite{Nogrette:2014fj}, it requires compensation of the non-flat surface of the SLM (``factory correction'') and the aberrations of the imaging system (aberration correction), as well as introducing an in-situ feedback system for the spot intensity. This work shows that these corrections can be obtained by solely using the SLM device in interferometric measurements. Additionally, we point out the general procedure of multiplexing the SLM for two wavelengths, 480$\,$nm and 780$\,$nm. 
\subsection{The spatial light modulator}\label{ch:SLM}
Aiming at high diffraction efficiency, we choose for a parallel-aligned nematic SLM, as this type offers a high fill factor of up to 90\% \cite{Bergamini:2004bl}. The fill factor refers to the active, voltage controlled part of the pixelated SLM surface. Any light hitting the SLM at the inter-pixel space leaves the device undeflected. Usually this part of the light is referred to as the ``zeroth order" deflection, and cannot be used for beam-steering purposes. Our SLM (HoloEye Photonics AG, PLUTO-BB HES 6010 BB) has a fill factor of 87\%, which yields a theoretical diffraction efficiency of $(\text{fill factor})^2=76\%$, which we find experimentally as well. The SLM is a phase-only modulator with $1920\times1080$ pixels, operating in the reflective mode, such that the light passes the active area twice and the modulation depth is twice as large as for transmission. The working principle is based on electrically changing the optical properties of an anisotropic liquid-crystal \cite{Huignard:1987go}. By controlling the applied voltage, the molecular alignment can be altered from parallel to perpendicular with reference to the display \cite{Banyal:2010fz,Neff:1990fg}. This rearrangement leads to a change in the refractive index and introduces a phase-shift on the incident light. As the voltage of each pixel can be modified independently, the SLM imprints a two-dimensional phase pattern $\phi(x,y)$ onto the reflected light. The modulation depth, meaning the maximal achievable phase shift, is usually a few multiples of $2\pi$. Therefore, phase information $\phi(x,y)$ is usually sent as $\phi(x,y)$ mod $2\pi$ to the device. \par
To ensure optimal phase modulation with minimal amplitude modulation, the polarization of the incoming light has to be aligned with the crystal axis. An advantage of the parallel-aligned nematic SLM is that they do not alter the polarization state of the reflected light, which is indispensable in quantum optics experiments.
\begin{figure}
\includegraphics{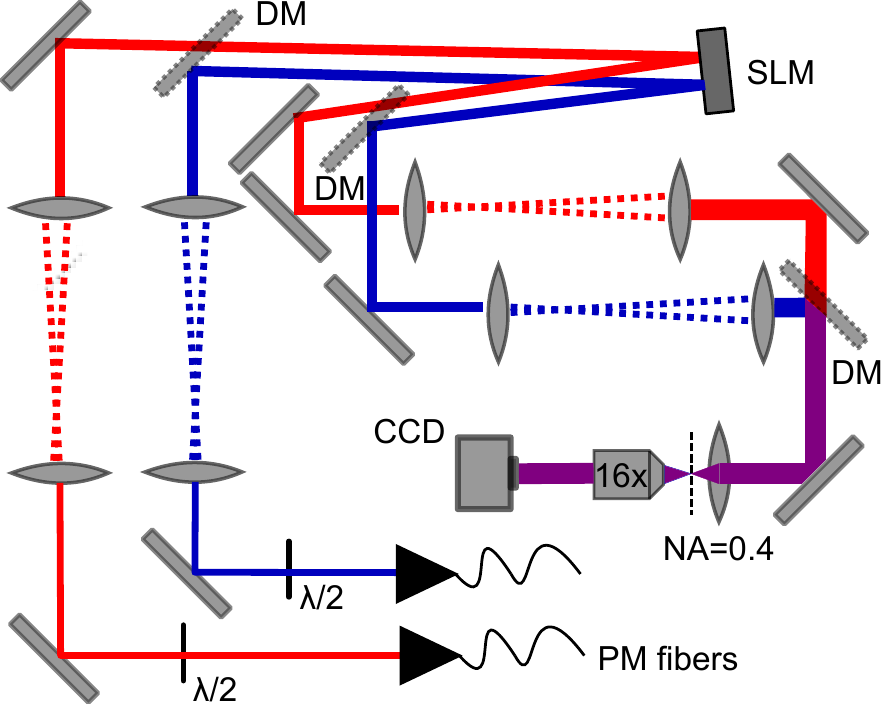}
\centering
\caption{Optical setup of the spatial light modulator (SLM, DM: dichroic mirror). Laser light at 480$\,$nm (blue) and 780$\,$nm (red) are coupled out of an optical fiber and expanded to match half the SLM screen in size. After leaving the SLM, the beams are separated at a DM, and enlarged by another factor of two, such that they match the diameter of the high NA imaging lens. The beams are overlapped again at another DM and focused by the lens. The intensity distribution in the focal plane is magnified by a factor of 16 using a microscope objective and imaged on a CCD camera.}
\label{fig:SLM_setup}
\end{figure}
\subsubsection{The optical setup}\label{ch:SLMOpt}
We require our SLM setup to steer both the 480$\,$nm and the 780$\,$nm light beams independently, both for ground state transitions and two-photon excitation of ${}^{87}$Rb to Rydberg states (compare section \ref{ch:ground} and section \ref{ch:rydberg}). Using two wavelengths at the same time adds additional complexity, first and foremost because the phase of the SLM pixels strongly depends on the wavelength. In addition, the optical components have refractive index dispersion which must be compensated. The atom chip experiment \cite{Leung:2014gw} features an imaging lens of high numerical aperture (NA=0.4), which is able to resolve individual traps in a $10\,\mu$m lattice. In our spatial addressing scheme this lens serves to transfer the SLM phase pattern to a desired intensity distribution at the trapped atoms' location. We built a test setup with an identical imaging lens, as seen in Fig. \ref{fig:SLM_setup}. Both wavelengths are coupled out of optical fibers to achieve a Gaussian beam profile. Subsequently they are expanded with a telescope, such that their diameter matches the height of the SLM display (the display size is 15.36$\,$mm$\times 8.64\,$mm). This maximizes the amount of available SLM pixels per beam. The beams are aligned next to each other by means of custom-made dichroic mirrors and guided such that each hits the center of one half of the SLM. Being reflected off of the SLM, the beams are separated using another dichroic mirror, and expanded again by a second telescope. The beams are overlapped by another dichroic mirror and focused by the lens. The beam sizes match the diameter of the imaging lens, thus, we exploit the full NA of this lens. The intensity distribution in the focal plane is magnified by a factor of 16 by means of a microscope objective, and imaged with a CCD camera. This extra magnification is required to gain sufficient spatial resolution of the intensity distribution, as the diffraction limited spot size in the focal plane is less than the camera pixel size.\par
The wavelengths are split into independent beam paths, as ray tracing simulations show that even with the use of achromatic optical elements, diffraction limited spot patterns in the focal plane cannot be achieved in a single beam path due to residual wavelength dispersion of the optical components. Independent optical components also allow for fine tuning. For example, the spacing of the telescope lenses can compensate for the difference in the imaging lens' focal length at 480$\,$nm and 780$\,$nm. The second telescope, in between the SLM and the imaging lens, is operated in a 4f-configuration. This means that the spacing between SLM and the first telescope lens, and between the second telescope lens and the imaging lens, has to match the respective focal lengths. As pointed out in \cite{Nogrette:2014fj}, this configuration ensures that the deflected beam hits the center of the imaging lens independent of the deflection angle at the SLM. We found earlier that without the 4f-configuration, we could only achieve small displacements in the focal plane without losing efficiency, especially for large beam sizes (compared to the lens) and long beam paths. The size of the telescope lenses and mirrors have been chosen such that we can achieve displacements of up to 300$\,\mu$m in the focal plane.\par
The angle between the incoming and reflected beam at the SLM was chosen to be $12^{\circ}$, which ensures that the beams can be separated at acceptable distances from the SLM, as well as that the angle of incidence at the SLM is around $6^{\circ}$. The latter is the optimal angle for the performance of our SLM. In general the diffraction efficiency of SLMs strongly depends on the angle of incidence of the incoming beam \cite{Lizana:2009ia}. To further maximize the diffraction efficiency, we introduce a $\lambda/2$-waveplate to align the light polarization with the long axis of the SLM display.
\subsubsection{Linearization of phase response}\label{ch:phaseResp}
\begin{figure}
\includegraphics{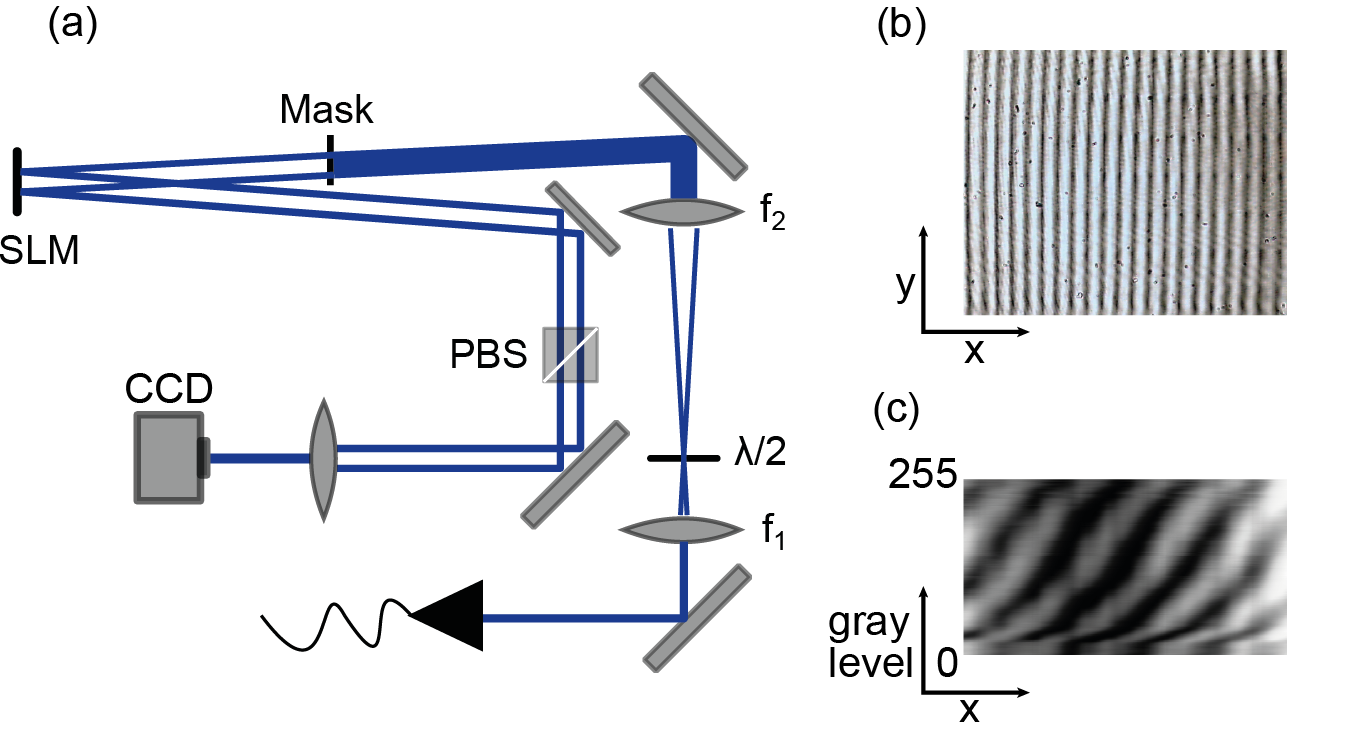}
\centering
\caption{(a) Mach-Zehnder interferometer as used to linearize the phase response of the SLM for both 480$\,$nm and 780$\,$nm. The laser light is expanded to fill two holes in a mask, whose spacing matches half the SLM horizontal length. The resultant beams hit one half of the SLM each, get reflected in zeroth order, pass through a polarizing beam splitter (PBS) and are overlapped with a lens onto a CCD camera. (b) Typical interference created at the camera. (c) Response of one horizontal line taken from the interference pattern to a change in gray level on the SLM}
\label{fig:mach}
\end{figure}
As mentioned before, the SLM is operated in a mod $2\pi$-mode, meaning that the maximum phase shift is $2\pi$. The phase shift $\phi(x,y)$ of every pixel is digitally programmed by an 8-bit value, called gray level, resulting in 256 different phase values. Ideally, the resultant phase values are equally spaced increments of $2\pi/256$. In practice, we need to calibrate the relationship between the gray levels and the voltage which is applied to the nematic crystals in order to achieve a linear phase response up to $2\pi$ \cite{Banyal:2010fz}. This calibration curve can be sent to the SLM, and serves as an interface between digital and phase information.\par
\begin{figure}
\includegraphics{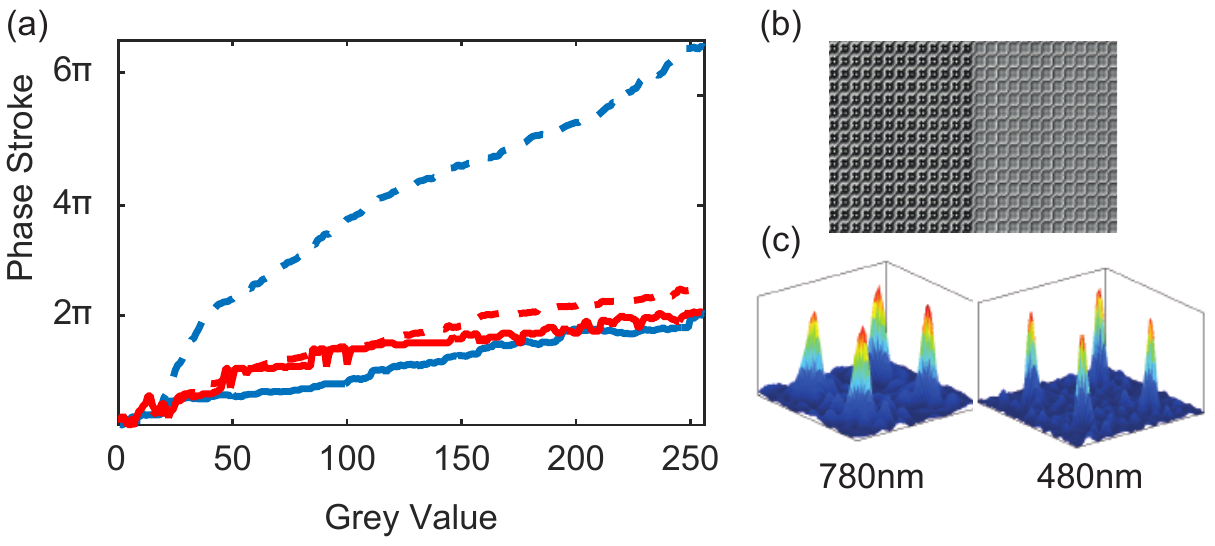}
\centering
\caption{(a) Phase response before (dashed line) and after the linearization (solid line) of the SLM response. (b) SLM phase pattern to create a 2x2 spot pattern compensating for the difference in phase response between 480$\,$nm and 780$\,$nm light. (c) Resultant normalized intensity pattern in the focal plane of the high NA lens with a 10$\,\mu$m spacing between the spots.}
\label{fig:double_spot}
\end{figure}

In order to calibrate the phase response we have to measure the actual phase value on the SLM. A simple approach is to use an interferometric measurement similar to a Mach-Zehnder interferometer, depicted in Fig. \ref{fig:mach} (a). We expand a laser beam out-coupled from an optical fiber at either 480$\,$nm or 780$\,$nm respectively, such that it fills a mask with two identical holes. The mask divides the beam into two independent parts, each of which hits the center of one half of the SLM. Afterwards, the two beams which are reflected from the backplane of the SLM are focused down, making them overlap and interfere on a CCD-camera. In Fig. \ref{fig:mach} (b) we show a typical interference pattern resulting from optical path length differences between the beam paths. These patterns are sensitive to thermal fluctuations and vibrations, inducing a shift of the interference fringes. In order to measure the relative phase value corresponding to a gray level, we change the gray level on one half of the SLM, leaving the other half unaltered. When ramping the gray levels from 0 to 255, the interference pattern is shifted, as can be seen in Fig. \ref{fig:mach} (c). The translation $x$ of the phase pattern is compared to the average spacing of the fringes $\bar{X}$, to give the SLM phase as $\phi=2\pi\times x/\bar{X}$. We take the spatial average of one fringe in the $x$-direction and plot the corresponding value against the applied gray value (see Fig. \ref{fig:double_spot} (a)). The measurements are performed with the default factory curve, and both show non-linear behavior and a phase shift of more than $2\pi$. There is a clear difference between the total phase stroke for 480$\,$nm and 780$\,$nm, which is expected as the ratio in phase shift should scale as the ratio 480$\,$nm/780$\,$nm. We use the information from the phase curves to change the gray value calibration curve to yield a linear response. A second measurement (Fig. \ref{fig:double_spot} (a)) shows the respective phase curves after applying the new calibration for 480$\,$nm and 780$\,$nm. The total phase shift matches $2\pi$, and the phase curve behaves more linearly. We are limited in our precision by the width of the fringes and the fluctuations mentioned above. In practice the applied calibration curves prove sufficient to produce high resolution spot patterns in the focal plane of Fig. \ref{fig:SLM_setup}.\par

As mentioned in section \ref{ch:SLMOpt}, we use the SLM for 480$\,$nm and 780$\,$nm laser light at the same time. However, only a single phase calibration curve can be introduced to the SLM, such that the maximal phase shift for one wavelength does not equal $2\pi$. In order to compensate for that, we change the gray values on one half of the SLM using the control software. If we use the calibration curve for 780$\,$nm, we naturally achieve a $2\pi$ phase shift for that wavelength and a larger one for 480$\,$nm. We rescale the gray values $g$ on the right half of the SLM, where the blue laser is incident, by $\widetilde{g}=\alpha\cdot g+g_\text{offset}$. In this formula, $\alpha$ and $g_\text{offset}$ are free parameters, which we optimize empirically by the resultant intensity patterns for 480$\,$nm. One would assume $\alpha=\frac{480}{780}$ as the optimal solution, but it proves to be $\alpha=0.4$. We show for the optimal parameters the phase pattern on the SLM for a 2x2 spot pattern with $10\,\mu$m spacing in Fig. \ref{fig:double_spot} (b). This pattern yields the intensity distribution shown in Fig. \ref{fig:double_spot} (c) in the focal plane of the high NA lens. For both wavelengths we achieve a spot pattern with well defined spacings of $10\,\mu$m. The difference in wavelengths is apparent in the different resolution limited spot sizes.
\subsubsection{Correction for the backplane non-flatness}\label{ch:Fac}
Due to the fabrication process the backplane of the SLM is not flat \cite{Jesacher:2007cv}, with measured deviations up to $3\lambda$ \cite{García-Márquez:2011jh}. As the phase modulation depth of the SLM is $2\pi$, the curvature of the backplane is sufficient to degrade the image quality in the focal plane. Hence we need to compensate for that deviation by introducing a corrective phase pattern on the SLM. To derive the spatial coordinates of a surface from the phase patterns of interferometric measurements using that same surface is a well know problem \cite{Hu:2007cf}. A common, multi-image solution is to use a modification of the original Le Carr\'e algorithm \cite{Carré:1966gy}, the so called phase-shifting interferometry (PSI) \cite{Magalhaes:2010jk}. In PSI, an interference image is created from two independent surfaces, one of which is known to be optically flat. Then the relative phase of the light coming from both surfaces is changed, e.g. by varying the optical path length, inducing changes in the interference pattern. Measuring that pattern for four different known phase values allows for retrieval of the phase deviation of the non-flat surface. In our approach we follow the methods presented in \cite{Amézquita:2011bd} in order to apply PSI to our system.\par
\begin{figure}
\includegraphics{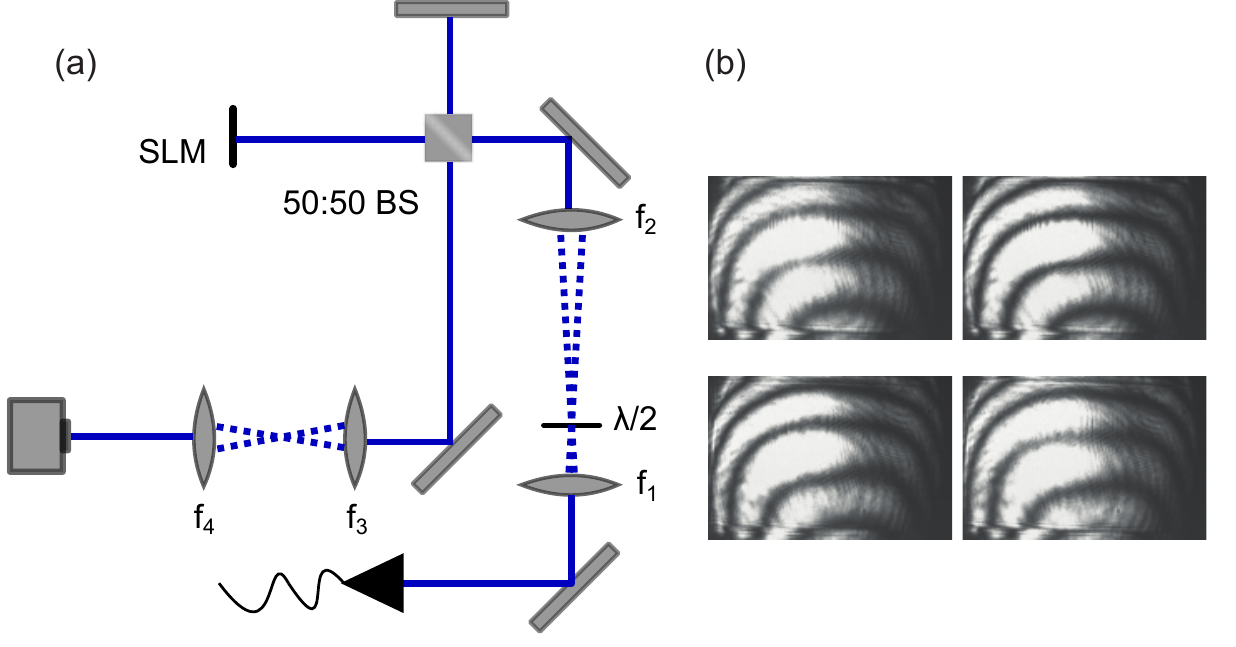}
\centering
\caption{(a) Michelson type interferometer involving the interference between a mirror and the SLM surface. The light is split and overlapped at a non-polarizing 50/50 beam splitter (BS) and projected with a telescope ($f_3$ and $f_4$) onto a CCD camera. (b) Four different images obtained for the relative phase values $0$, $\pi/2$, $\pi$ and $3\pi/2$ in phase shifting interferometry.}
\label{fig:michelson}
\end{figure}
The test setup, shown in Fig. \ref{fig:michelson} (a), is very similar to a Michelson-interferometer. The optically flat reference surface is implemented by a dielectric mirror with a surface flatness of better than $\lambda/10$. The incident laser light at 780$\,$nm is expanded and split by a 50/50 non-polarizing beam splitter into two arms, filling most of the SLM backplane and the mirror. The light is reflected in zeroth order from the SLM, and partially overlapped at the same beam splitter with the light originating from the mirror. The overlapping beams are projected onto a CCD camera such that the size of the interference pattern matches the size of the camera chip. The change in relative phase is achieved by introducing a phase shift on the SLM. As the interference pattern is sensitive to environmental influences, we swiftly ramp the phase on the SLM from 0 to $2\pi$, triggering the camera to take four images during the ramp yielding interference pattern at a phase of $0$, $\pi/2$, $\pi$ and $3\pi/2$, shown in Fig. \ref{fig:michelson} (b). We optimize the interference pattern such that it shows the minimal number of fringes. Then we can assume normal incidence on both surfaces which simplifies the mathematical analysis. \par
\begin{figure}
\includegraphics{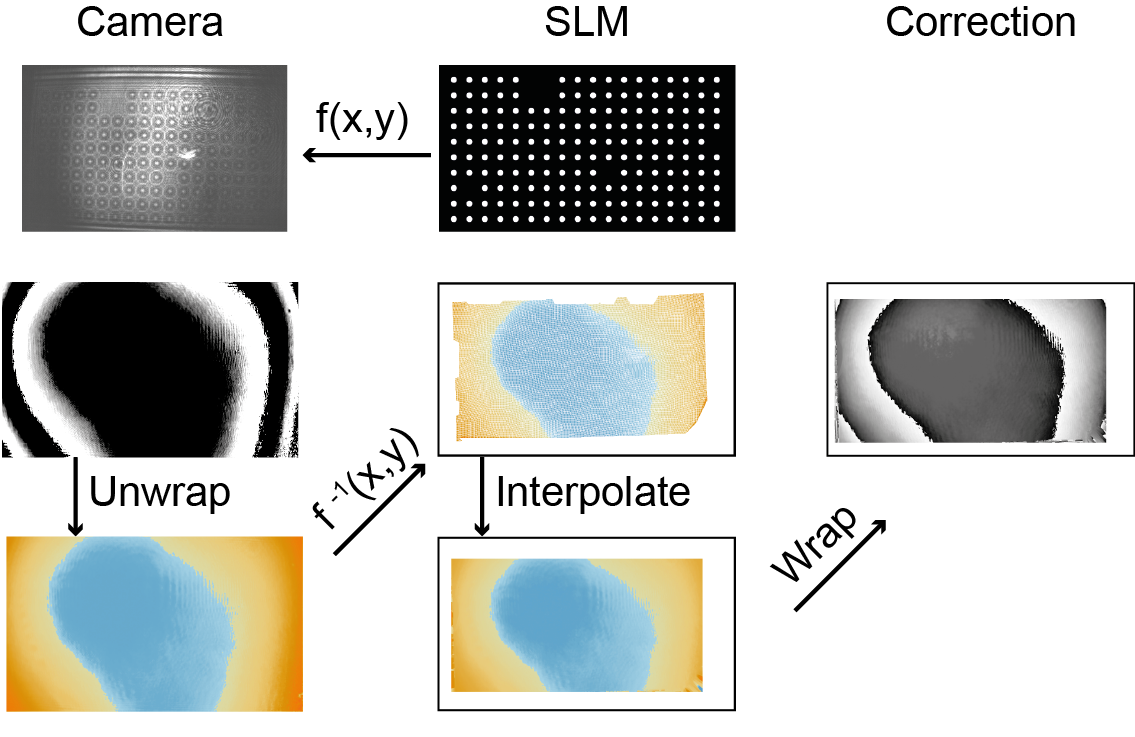}
\centering
\caption{The principle of the factory correction, showing the images and phases on the camera and in the SLM plane. The coordinate transformation $f(x,y)$ between camera and SLM plane is found by introducing a pattern of regularly spaced spots with some voids on the SLM, and imaging the corresponding intensity pattern on the camera. The phase found by PSI at the camera has to be unwrapped before transforming the values to the SLM plane. The transformed values have to be interpolated, accounting for the lower resolution of the camera. The resulting phase pattern has to be wrapped mod $2\pi$. The inverse of that is used as the factory correction.}
\label{fig:wrapping_sum}
\end{figure}

The resultant four images contain a two-dimensional array of intensity information $I_i(x,y)$ with $i\in \{1,2,3,4\}$ in the camera plane (x and y denote the index of the relative camera pixel), from which we want to extract a similar array of phase values $\phi(x,y)$. For the four step algorithm it can be shown that images taken with the relative phase of $\delta_i=(i-1)\cdot \frac{\pi}{2}$, yield the following relation \cite{Malacara:2007tk}:
\begin{align*}
I_1(x,y)&=I_0(x,y)+I_\text{fr}(x,y)\cos[\phi(x,y)]\\
I_2(x,y)&=I_0(x,y)+I_\text{fr}(x,y)\cos[\phi(x,y)+\pi/2]\\
I_3(x,y)&=I_0(x,y)+I_\text{fr}(x,y)\cos[\phi(x,y)+\pi]\\
I_4(x,y)&=I_0(x,y)+I_\text{fr}(x,y)\cos[\phi(x,y)+3\pi/2],\\
\end{align*}
where $I_0$ is the summed individual beam intensity and $I_\text{fr}$ is the fringe modulation. These are four equations per pixel for the three unknowns $I_0$, $I_\text{fr}$ and $\phi$, which can be solved for $\phi(x,y)$:
\[
\frac{I_4-I_2}{I_1-I_3}=\frac{\sin[\phi(x,y)]}{\cos[\phi(x,y)]}=\tan[\phi(x,y)].
\]
We can use the inverse tangent with two arguments, as the sign of numerator and denominator are known, to retrieve $\phi(x,y)$ in the range $[-\pi,\pi)$ . Now the phase values are known mod $2\pi$ in the camera plane (see Fig. \ref{fig:wrapping_sum}), which in principle would be sufficient as the SLM phase is encoded in the same fashion. However, we need to relate the measured phase values in the camera plane to the relative phase values in the SLM plane. This relation is given by a transformation describing the imaging properties of our optical system. We first have to retrieve the actual phase from the values mod $2\pi$, because this phase is more robust to the applied transformation. Retrieving the actual phase is a non-trivial problem in more than two dimensions \cite{Malacara:2007tk}, usually referred to as phase unwrapping. A multitude of different algorithms have been developed to minimize errors and mathematical artifacts \cite{Huntley:1989kt,Buckland:1995ko} in the phase unwrapping.
We unwrap the measured phase array $\phi(x,y)$ and retrieve a smooth two-dimensional phase pattern $\Phi(x,y)$ with values up to $4\pi$, consistent with the previously mentioned literature values. We are left with transforming the phase values $\Phi(x,y)$ to the corresponding values $\Phi^\prime(x^\prime,y^\prime)$ in the SLM plane. Therefore, we need to measure the coordinate transformation $f(x,y)$ between the SLM $(x^\prime,y^\prime)$ and the camera (x,y) coordinates. We introduce a pattern of regularly spaced spots (with some defects for better reference, see Fig. \ref{fig:wrapping_sum}) onto the SLM, and observe the resulting intensity pattern in the camera plane.
Every spot is clearly identifiable besides those which are not captured by the imaging system or outside of the laser intensity profile. As there is now a one-to-one correspondence between spots in the camera and SLM plane, we can retrieve the desired transformation function $f$ for the spatial coordinates and apply $\Phi^\prime(x^\prime,y^\prime)=\Phi(f^{-1}(x^\prime,y^\prime))$. The resultant phase pattern does not fill the whole SLM canvas, as not all spots are imaged on the camera. Furthermore, there are missing pixel values as the resolution of the camera is smaller than the SLM resolution. We fill those missing elements by interpolating between neighbouring pixel values. The resultant phase $\Phi^\prime(x^\prime,y^\prime)$ is mapped by mod $2\pi$ to values $\phi^\prime(x^\prime,y^\prime)$, from which we retrieve the correction sent to the SLM as $2\pi-\phi^\prime$. This correction is usually called factory correction, as it accounts for the non-flatness of the surface due to the manufacturing process.\par

\begin{figure}
\includegraphics{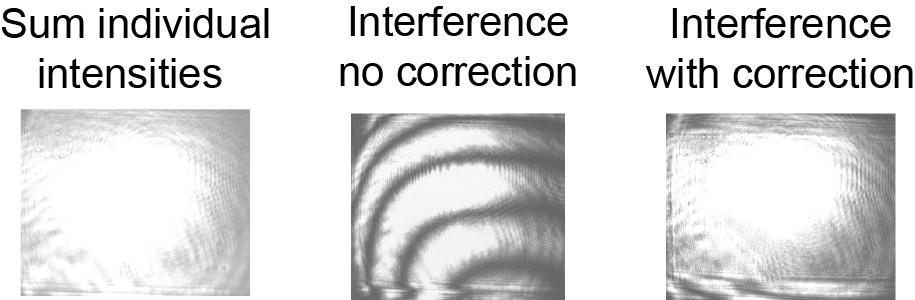}
\centering
\caption{Influence of the backplane curvature correction on the outcome of the phase-shifting interferometry.}
\label{fig:wrapping_res}
\end{figure}

\subsubsection{Aberration correction}\label{ch:aber}
Optical aberrations are an unwanted feature of an optical system, degrading its imaging quality. In practice, every optical system suffers from a variety of different aberrations, which have to be determined and compensated for. There is a multitude of ways to measure aberrations in an optical system \cite{Malacara:2007tk}, and many measures can be taken to improve the performance. For example, one can compensate for chromatic aberrations by introducing achromatic optical elements, or for spherical aberrations by using aspheric lenses. Besides that, it is common to use adaptive optics as an active compensation method, for example in astronomy \cite{Hardy:2000kk} or biology \cite{Kubby:2013fu}. This is especially useful if there are time-varying aberrations present, like atmospheric turbulence in astronomy, or if the aberrations are complex. A standard technical device for aberration analysis is the Shack-Hartmann (SH) sensor \cite{Shack:1971vy}, which is often used with adaptive optics \cite{PEARSON:1979kf}. As an example, SH sensors are used in astronomy for telescopes \cite{Fugate:1994kx}, or in biology to measure the aberration of the human eye \cite{Liang:1994dp}. The SH sensor consists of a multi-lenslet array, which images the incoming wavefront to an array of single spots on a CCD camera. The basic working idea is to use the measured displacement in the imaging plane to estimate the local wavefront tilt incident on an individual lenslet. The SH sensor is calibrated with a near-perfect plane wave to localize the optimal spot position on the CCD camera. Any wavefront with a local phase gradient will lead to a deviation from that optimal position.\par
\begin{figure}
\includegraphics{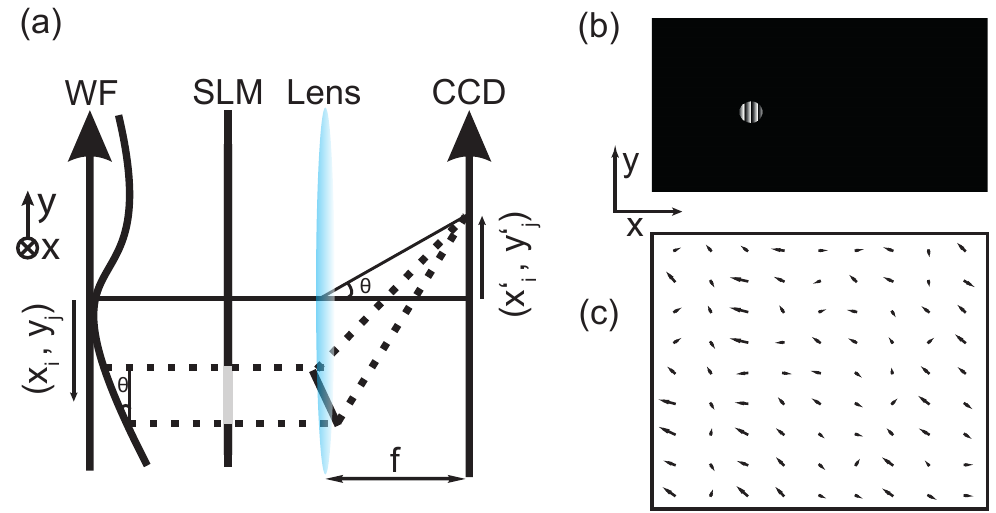}
\centering
\caption{(a) Concept of the SLM based Shack-Hartmann sensor. On the SLM we place an aperture with diameter $500\,\mu$m positioned at $(x_i,y_j)$.The aperture contains a linear grating sufficient to steer the light inside that aperture away from the 0th order of the SLM. The local wavefront gradient is transformed by the lens of the imaging system into a translation $(x^\prime_i,y^\prime_j)$ in the focal plane. (b) The SLM pattern (zoom in) corresponding to a single measurement at a given position $(x_i,y_j)$. (c) Part of the resultant pattern of translation vectors $(x^\prime_i,y^\prime_j)$ plotted at the respective position $(x_i,y_j)$ in the SLM plane for the 780$\,$nm light. The area available for the measurement is $4\times4\,$mm in size corresponding to most of the left half of the SLM dedicated to the 780$\,$nm light.}
\label{fig:SH}
\end{figure}

Naturally, a SLM qualifies as a candidate for adaptive optics, as we can adjust the local phase of an incident wavefront with high resolution. Additionally, we want to use the SLM as a sensor for optical aberrations of our imaging system without necessitating an additional commercial device. Our scheme mimics the working principle of the SH sensor, and resembles the one presented in \cite{Bowman:2010dq}. We aim primarily at improving the quality of spot patterns in the focal plane of our high NA lens, which can be achieved by compensating aberrations with the SLM itself \cite{Nogrette:2014fj}. The idea behind our method is depicted in Fig. \ref{fig:SH} (a). If light  propagates in the $z$-direction, it can be described at a given point $z=z_0$ by the wavefront $\Psi(x,y)$. For a perfect plane wave, the wavefront is flat, hence $\frac{\partial \Psi(x_0,y_0)}{\partial x}=\frac{\partial \Psi(x_0,y_0)}{\partial y}=0$ at any point $(x_0,y_0)$.
Optical aberrations in the beam path will influence the wavefront, introducing non-vanishing gradients. Locally the wavefront can be thought of as tilted by an angle $\theta$ with respect to the plane perpendicular to the $z$-direction. Such a tilt leads to a displacement from the focal point of a lens imaging that wavefront. In our setup we introduce a pattern like shown in Fig. \ref{fig:SH} (b) onto the SLM. The pattern consists of a spot with diameter $500\,\mu$m, containing a linear grating which steers the light away from the zeroth order of the SLM. The rest of the light coincides with the zeroth order and is blocked in the focus of the second telescope in Fig.~\ref{fig:SLM_setup}. Hence the spot acts like a local aperture, selecting only a small part of the wavefront. We image the deflected light on our CCD camera, and measure the displacement in the focal plane compared to the image of the complete wavefront. All measurements are performed with the factory correction from section \ref{ch:Fac} present on the SLM. We move the spot around the SLM canvas in steps of $250\,\mu$m in $x$- and $y$-direction, thus super-sampling the wavefront. Any change in the wavefront smaller than the diameter of the spot cannot be distinguished. In order to increase the resolution of this method, we need to decrease the spot size. However, there is a limit to the spot size due to two reasons: the intensity of light coming from smaller spots is too low to be detected by the CCD, and smaller spots mean an effective decrease in NA of the imaging lens, thus leading to larger images which increase the uncertainty in the image center position. Using the mentioned spacing, we obtain a displacement vector $(x^\prime_i,y^\prime_j)$ in the focal plane for every spot position $(x_i,y_j)$ on the SLM (see Fig. \ref{fig:SH} (c)) from fitting a 2D-Gaussian function to the image. The data is limited to a 4 x 4$\,$mm area around the 780$\,$nm beam center on the SLM due to the finite beam size.\par     
There is a simple geometric relation between the gradient of the wavefront $\Psi$ at $(x_i,y_j)$ and the measured displacement $(x_i^\prime,y_j^\prime)$:
\begin{align*}
\frac{\partial \Psi(x_i,y_j)}{\partial x}&= \frac{x^\prime_i}{f^\prime}\\
\frac{\partial \Psi(x_i,y_j)}{\partial y}&= \frac{y^\prime_j}{f^\prime},
\end{align*}
where $f^\prime=m\cdot f$ is the modified focal length of our imaging system, with $m$ being the magnification of the second telescope.
\begin{figure}
\includegraphics{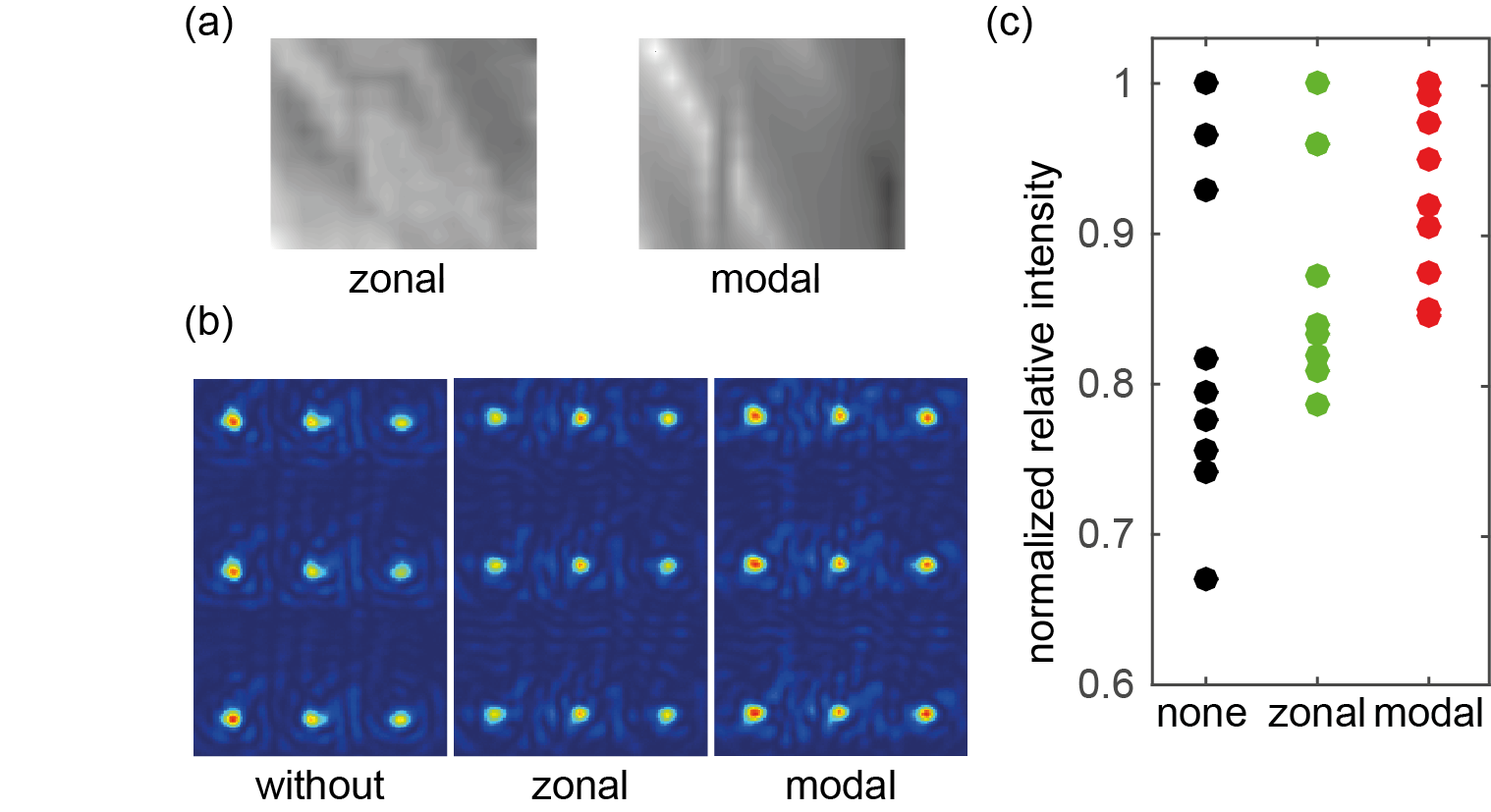}
\centering
\caption{(a) The correction pattern imposed on the SLM as obtained from the zonal and modal method. (b) Resultant 3x3 spot pattern with a $10\,\mu$m spacing in the horizontal direction obtained in the focal plane for applying no correction, or using the zonal or modal correction. (c) The normalized intensity in a 700$\,$nm x 700$\,$nm area around the spot maximum of all spots depicted in (b), using no correction, the zonal or the modal correction.}
\label{fig:spots}
\end{figure}
From our measurements, we receive an array of estimators for the local gradient of the wavefront in the $x$- and the $y$-direction. We now need to retrieve the actual wavefront from those gradients. This is a well known problem, for which two methods have been developed, called the zonal and modal reconstruction methods \cite{Southwell:1980fc}. The zonal method reconstructs the wavefront from the gradients in every measurement point, whereas the modal method expresses the wavefront as a superposition of basis functions, and tries to fit the gradients to the data. Both methods have advantages and disadvantages depending on the specific application. Therefore we use both methods in our case and evaluate which one performs better. Mathematically, we use the approach depicted in \cite{Topa:2002dt} based on the singular value decomposition (SVD). The zonal approach necessitates a set of basis functions, for which one normally uses Zernike polynomials, as they allow a direct interpretation in terms of actual optical aberrations \cite{Malacara:2007tk}. As Zernike polynomials are defined on the unit sphere, they have to be transformed to be pairwise orthogonal in a rectangular geometry, matching our measurement. We take the first 15 rectangular Zernike polynomials as defined in \cite{Mahajan:2007ko}.\par
We convert the wavefront $\Psi$ obtained from both methods to a phase information $\phi$ according to \[\phi(x,y)=\frac{2\pi\cdot \Psi(x,y)}{\lambda},\] and use the inverse of that for our correction on the SLM. The resultant phase pattern for both the modal and zonal method are shown in Fig. \ref{fig:spots} (a). As the zonal method yields a set of discrete values with $250\,\mu$m spacing, the zonal phase pattern is interpolated. There are visible differences between both patterns, as well as structural similarities. Naturally, the modal phase pattern is smoother due to the underlying functional description. Both patterns do not allow for an easy explanation in terms of standard aberrations. However, the term corresponding to a tilt in the $x$-direction is the largest coefficient in the Zernike decomposition. To validate that the correctional phase patterns compensate for the aberrations in our optical system, we look at intensity spot patterns created in the focal plane. Ideally those spots are resolution limited and possess identical intensity and shape. The spot patterns are shown for the example of a three by three pattern with a $10\,\mu$m x $20\,\mu$m spacing for the case of no correction, the zonal and the modal correction in Fig. \ref{fig:spots} (b). Visually, the spots become more uniform in intensity and shape. The spot pattern without correction shows a strong decrease of intensity in the $x$-direction, which is related to the tilt in that direction predicted from the zonal method. To better quantify the performance we define an easy figure of merit by the summed intensity in an area of 700$\,$nm x 700$\,$nm around each spot center, roughly corresponding to the atom cloud size in our magnetic microtraps. The results are normalized to the highest summed intensity in every pattern, and plotted in Fig. \ref{fig:spots} (c), which show that there is a spread of up to 33\% in the uncorrected case. This spread in intensity is reduced by both correction patterns, with the modal correction performing better than the zonal correction. This can be most likely related to the fact that the modal method compensates better for the finite resolution of our SH method.\par
In conclusion, we showed that we can use the SLM for both analyzing and correcting optical aberrations in our system. The detected aberrations can be related to the SLM itself, as for the non-flatness of the backplane, or to the optical components in the rest of the imaging system. The backplane curvature can be corrected to yield an effective SLM surface with a flatness of around $\lambda/10$, whereas the modal correction pattern significantly improves the imaging performance. Both the factory and modal correction are required on the SLM. However, with all corrections present, we still have a spread in intensity of around 15\% in our spot patterns. Furthermore, the performance decreases if the amount of spots is increased, as the finite resolution of the SH methods becomes more influential. To achieve more uniform spot patterns we have to involve an active feedback system, as was previously found in \cite{Nogrette:2014fj}.

%% file: OptTech_min.bbl
\begin{thebibliography}{100}

\bibitem{Saffman:2010du}
M.~Saffman, T.~G. Walker, and K.~M{\o}lmer, ``{Quantum information with Rydberg
  atoms},'' {\em Reviews of Modern Physics}, vol.~82, no.~3, p.~2313, 2010.

\bibitem{Cote:2001uo}
R.~Cote, L.~M. Duan, D.~Jaksch, J.~I. Cirac, and P.~Zoller, ``{Dipole blockade
  and quantum information processing in mesoscopic atomic ensembles},'' {\em
  Physical Review Letters}, 2001.

\bibitem{Weimer:2010vq}
H.~Weimer, M.~M{\"u}ller, I.~Lesanovsky, P.~Zoller, and H.~P. B{\"u}chler, ``{A
  Rydberg quantum simulator},'' {\em Nature Physics}, vol.~6, no.~5, p.~382,
  2010.

\bibitem{Webster:2013tr}
S.~C. Webster, S.~Weidt, K.~Lake, J.~J. McLoughlin, and W.~K. Hensinger,
  ``{Simple Manipulation of a Microwave Dressed-State Ion Qubit},'' {\em
  Physical Review Letters}, vol.~111, no.~14, 2013.

\bibitem{Treutlein:2004ft}
P.~Treutlein, P.~Hommelhoff, T.~Steinmetz, T.~W. H{\"a}nsch, and J.~Reichel,
  ``{Coherence in Microchip Traps},'' {\em Physical Review Letters}, vol.~92,
  no.~20, 2004.

\bibitem{Gallagher:1994hu}
T.~F. Gallagher, {\em {Rydberg Atoms}}.
\newblock Cambridge Univsersity Press, 1994.

\bibitem{Dudin:2012hm}
Y.~O. Dudin and A.~Kuzmich, ``{Strongly Interacting Rydberg Excitations of a
  Cold Atomic Gas},'' {\em Science}, vol.~336, no.~6083, p.~887, 2012.

\bibitem{Nipper:2012jv}
J.~Nipper, J.~B. Balewski, A.~T. Krupp, B.~Butscher, R.~L{\"o}w, and T.~Pfau,
  ``{Highly Resolved Measurements of Stark-Tuned F{\"o}rster Resonances between
  Rydberg Atoms},'' {\em Physical Review Letters}, vol.~108, no.~11, 2012.

\bibitem{Vogt:2006kw}
T.~Vogt, M.~Viteau, J.~Zhao, A.~Chotia, D.~Comparat, and P.~Pillet, ``{Dipole
  Blockade at F{\"o}rster Resonances in High Resolution Laser Excitation of
  Rydberg States of Cesium Atoms},'' {\em Physical Review Letters}, vol.~97,
  no.~8, 2006.

\bibitem{He:1990dp}
X.~He, B.~Li, A.~Chen, and C.~Zhang, ``{Model-potential calculation of
  lifetimes of Rydberg states of alkali atoms},'' {\em Journal of Physics B:
  Atomic, Molecular and Optical Physics}, vol.~23, no.~4, p.~661, 1990.

\bibitem{Gaëtan:2009fq}
A.~Ga{\"e}tan, Y.~Miroshnychenko, T.~Wilk, A.~Chotia, M.~Viteau, D.~Comparat,
  P.~Pillet, A.~Browaeys, and P.~Grangier, ``{Observation of collective
  excitation of two individual atoms in the Rydberg blockade regime},'' {\em
  Nature Physics}, vol.~5, no.~2, p.~115, 2009.

\bibitem{Zhang:2010gk}
X.~L. Zhang, L.~Isenhower, A.~T. Gill, T.~G. Walker, and M.~Saffman,
  ``{Deterministic entanglement of two neutral atoms via Rydberg blockade},''
  {\em Physical Review A}, vol.~82, no.~3, 2010.

\bibitem{Isenhower:2010uq}
E.~Urban, T.~Henage, and T.~A. Johnson, ``{Demonstration of a Neutral Atom
  Controlled-NOT Quantum Gate},'' {\em Physical Review Letters}, vol.~104,
  no.~1, 2010.

\bibitem{Schauß:2012ee}
P.~Schau{\ss}, M.~Cheneau, M.~Endres, T.~Fukuhara, S.~Hild, A.~Omran, T.~Pohl,
  C.~Gross, S.~Kuhr, and I.~Bloch, ``{Observation of spatially ordered
  structures in a two-dimensional Rydberg gas},'' {\em Nature}, vol.~491,
  no.~7422, p.~87, 2012.

\bibitem{Viteau:2011ik}
M.~Viteau, M.~G. Bason, J.~Radogostowicz, N.~Malossi, D.~Ciampini, O.~Morsch,
  and E.~Arimondo, ``{Rydberg Excitations in Bose-Einstein Condensates in
  Quasi-One-Dimensional Potentials and Optical Lattices},'' {\em Physical
  Review Letters}, vol.~107, no.~6, 2011.

\bibitem{Urban:2009jd}
E.~Urban, T.~A. Johnson, T.~Henage, L.~Isenhower, D.~D. Yavuz, T.~G. Walker,
  and M.~Saffman, ``{Observation of Rydberg blockade between two atoms},''
  {\em Nature Physics}, vol.~5, no.~2, p.~110, 2009.

\bibitem{Gunter:2013fv}
G.~Gunter, H.~Schempp, M.~Robert-de Saint-Vincent, V.~Gavryusev, S.~Helmrich,
  C.~S. Hofmann, S.~Whitlock, and M.~Weidem{\"u}ller, ``{Observing the Dynamics
  of Dipole-Mediated Energy Transport by Interaction-Enhanced Imaging},'' {\em
  Science}, vol.~342, no.~6161, p.~954, 2013.

\bibitem{Walker:2012bi}
T.~G. Walker and M.~Saffman, ``{Entanglement of Two Atoms Using Rydberg
  Blockade},'' {\em Advances In Atomic, Molecular, and Optical Physics},
  vol.~61, pp.~81--115, 2012.

\bibitem{Bize:1999gm}
S.~Bize, Y.~Sortais, M.~S. Santos, C.~Mandache, A.~Clairon, and C.~Salomon, ``{
  High-accuracy measurement of the 87 Rb ground-state hyperfine splitting in an
  atomic fountain },'' {\em Europhysics Letters (EPL)}, vol.~45, no.~5, p.~558,
  1999.

\bibitem{Steck:0vg}
D.~A. Steck, ``Rubidium 87 d line data.''
  \url{http://steck.us/alkalidata/rubidium87numbers.1.6.pdf}, 2001.

\bibitem{HAPPER:1972ed}
W.~HAPPER, ``{Optical Pumping},'' {\em Reviews of Modern Physics}, vol.~44,
  no.~2, p.~169, 1972.

\bibitem{Weitenberg:2011gn}
C.~Weitenberg, M.~Endres, J.~F. Sherson, M.~Cheneau, P.~Schau{\ss},
  T.~Fukuhara, I.~Bloch, and S.~Kuhr, ``{Single-spin addressing in an atomic
  Mott insulator},'' {\em Nature}, vol.~471, no.~7338, p.~319, 2011.

\bibitem{Yavuz:2006gj}
D.~D. Yavuz, P.~B. Kulatunga, E.~Urban, T.~A. Johnson, N.~Proite, T.~Henage,
  T.~G. Walker, and M.~Saffman, ``{Fast Ground State Manipulation of Neutral
  Atoms in Microscopic Optical Traps},'' {\em Physical Review Letters},
  vol.~96, no.~6, 2006.

\bibitem{Thomas:1982ev}
J.~E. Thomas, P.~R. Hemmer, S.~Ezekiel, C.~C. Leiby, R.~H. Picard, and C.~R.
  Willis, ``{Observation of Ramsey Fringes Using a Stimulated, Resonance Raman
  Transition in a Sodium Atomic Beam},'' {\em Physical Review Letters},
  vol.~48, no.~13, p.~867, 1982.

\bibitem{Feng:2014gz}
Y.~Feng, H.~Xue, X.~Wang, S.~Chen, and Z.~Zhou, ``{Observation of Ramsey
  fringes using stimulated Raman transitions in a laser-cooled continuous
  rubidium atomic beam},'' {\em Applied Physics B}, vol.~118, no.~1, p.~139,
  2014.

\bibitem{Enloe:1965it}
L.~H. Enloe and J.~L. Rodda, ``{Laser phase-locked loop},'' {\em Proceedings of
  the IEEE}, vol.~53, no.~2, p.~165, 1965.

\bibitem{Leeb:1982et}
W.~R. Leeb, ``{Frequency synchronization and phase locking of CO2 lasers},''
  {\em Applied Physics Letters}, vol.~41, no.~7, p.~592, 1982.

\bibitem{Steele:1983kq}
R.~C. Steele, ``{Optical phase-locked loop using semiconductor laser diodes},''
  {\em Electronics Letters}, vol.~19, no.~2, p.~69, 1983.

\bibitem{Prevedelli:1995tfbaca}
M.~Prevedelli and T.~Freegarde, ``{Phase locking of grating-tuned diode
  lasers},'' {\em Physics B Lasers}, 1995.

\bibitem{Höckel:2008ft}
D.~H{\"o}ckel, M.~Scholz, and O.~Benson, ``{A robust phase-locked diode laser
  system for EIT experiments in cesium},'' {\em Applied Physics B}, vol.~94,
  no.~3, p.~429, 2008.

\bibitem{Pearman:2002kb}
C.~P. Pearman, C.~S. Adams, S.~G. Cox, P.~F. Griffin, D.~A. Smith, and I.~G.
  Hughes, ``{Polarization spectroscopy of a closed atomic transition:
  applications to laser frequency locking},'' {\em Journal of Physics B:
  Atomic, Molecular and Optical Physics}, vol.~35, no.~24, p.~5141, 2002.

\bibitem{Borklund:1983vv}
G.~C. Borklund, M.~Levenson, W.~Lenth, and C.~Ortiz, ``{Frequency-modulation
  (FM) spectroscopy},'' {\em Appl. Phys. B, 1983}, 1983.

\bibitem{Ramos:1990fs}
R.~T. Ramos and A.~J. Seeds, ``{Delay, linewidth and bandwidth limitations in
  optical phase-locked loop design},'' {\em Electronics Letters}, vol.~26,
  no.~6, p.~389, 1990.

\bibitem{Mirabbasi:1999er}
S.~Mirabbasi and K.~Martin, ``{Design of loop filter in phase-locked loops},''
  {\em Electronics Letters}, vol.~35, no.~21, p.~1801, 1999.

\bibitem{Isenhower:2010ke}
L.~Isenhower, E.~Urban, X.~L. Zhang, A.~T. Gill, T.~Henage, T.~A. Johnson,
  T.~G. Walker, and M.~Saffman, ``{Demonstration of a Neutral Atom
  Controlled-NOT Quantum Gate},'' {\em Physical Review Letters}, vol.~104,
  no.~1, 2010.

\bibitem{Theodosiou:1984cp}
C.~E. Theodosiou, ``{Lifetimes of alkali-metal{\textemdash}atom Rydberg
  states},'' {\em Physical Review A}, vol.~30, no.~6, p.~2881, 1984.

\bibitem{Ryabtsev:2011cd}
I.~I. Ryabtsev, I.~I. Beterov, D.~B. Tretyakov, V.~M. Entin, and E.~A.
  Yakshina, ``{Doppler- and recoil-free laser excitation of Rydberg states via
  three-photon transitions},'' {\em Physical Review A}, vol.~84, no.~5, 2011.

\bibitem{Abel:2009ig}
R.~P. Abel, A.~K. Mohapatra, M.~G. Bason, J.~D. Pritchard, K.~J. Weatherill,
  U.~Raitzsch, and C.~S. Adams, ``{Laser frequency stabilization to excited
  state transitions using electromagnetically induced transparency in a cascade
  system},'' {\em Applied Physics Letters}, vol.~94, no.~7, p.~071107, 2009.

\bibitem{Ma:1994kx}
L.-S. Ma, P.~Jungner, J.~Ye, and J.~L. Hall, ``{Delivering the same optical
  frequency at two places: accurate cancellation of phase noise introduced by
  an optical fiber or other time-varying path},'' {\em Optics Letters},
  vol.~19, no.~21, p.~1777, 1994.

\bibitem{Burck:2002kv}
F.~Du~Burck and O.~Lopez, ``{Stabilisation of laser beam intensity at 2.5
  MHz},'' {\em Electronics Letters}, vol.~38, no.~23, p.~1447, 2002.

\bibitem{Takahashi:2008do}
K.~Takahashi, M.~Ando, and K.~Tsubono, ``{Stabilization of laser intensity and
  frequency using optical fiber},'' {\em Journal of Physics: Conference
  Series}, vol.~122, p.~012016, 2008.

\bibitem{Liu:2013ib}
F.~Liu, C.~Wang, L.~Li, and L.~Chen, ``{Long-term and wideband laser intensity
  stabilization with an electro-optic amplitude modulator},'' {\em Optics {\&}
  Laser Technology}, vol.~45, p.~775, 2013.

\bibitem{Ogata:2002tz}
K.~Ogata, {\em {Modern Control Engineering}}.
\newblock Pearson Education International, 2002.

\bibitem{Inaba:2004hm}
H.~Ina, T.~Ikegami, F.-L. Hong, A.~Onae, Y.~Koga, T.~R. Schibli, K.~Minoshima,
  H.~Matsumoto, S.~Yamadori, O.~Tohyama, and S.~I. Yamaguchi, ``{Phase locking
  of a continuous-wave optical parametric oscillator to an optical frequency
  comb for optical frequency synthesis},'' {\em IEEE Journal of Quantum
  Electronics}, vol.~40, no.~7, p.~929, 2004.

\bibitem{Schibli:2005km}
T.~R. Schibli, K.~Minoshima, F.~L. Hong, H.~Ina, Y.~Bitou, A.~Onae, and
  H.~Matsumoto, ``{Phase-locked widely tunable optical single-frequency
  generator based on a femtosecond comb},'' {\em Optics Letters}, vol.~30,
  no.~17, p.~2323, 2005.

\bibitem{Grimmel:2015hr}
J.~Grimmel, M.~Mack, F.~Karlewski, F.~Jessen, M.~Reinschmidt, N.~S{\'a}ndor,
  and J.~Fort{\'a}gh, ``{Measurement and numerical calculation of Rubidium
  Rydberg Stark spectra},'' {\em New Journal of Physics}, vol.~17, no.~5,
  p.~053005, 2015.

\bibitem{Bohlouli-Zanjani:2006hu}
P.~Bohlouli-Zanjani, K.~Afrousheh, and J.~D.~D. Martin, ``{Optical transfer
  cavity stabilization using current-modulated injection-locked diode
  lasers},'' {\em Review of Scientific Instruments}, vol.~77, no.~9, p.~093105,
  2006.

\bibitem{Schünemann:1999cv}
U.~Sch{\"u}nemann, H.~Engler, R.~Grimm, M.~Weidem{\"u}ller, and
  M.~Zielonkowski, ``{Simple scheme for tunable frequency offset locking of two
  lasers},'' {\em Review of Scientific Instruments}, vol.~70, no.~1, p.~242,
  1999.

\bibitem{Mogensen:1985gz}
F.~Mogensen, H.~Olesen, and G.~Jacobsen, ``{Locking conditions and stability
  properties for a semiconductor laser with external light injection},'' {\em
  IEEE Journal of Quantum Electronics}, vol.~21, no.~7, p.~784, 1985.

\bibitem{Paldus:2005es}
B.~A. Paldus and A.~A. Kachanov, ``{An historical overview of cavity-enhanced
  methods},'' {\em Canadian Journal of Physics}, vol.~83, no.~10, p.~975, 2005.

\bibitem{Dahmani:1987it}
B.~Dahmani, L.~Hollberg, and R.~Drullinger, ``{Frequency stabilization of
  semiconductor lasers by resonant optical feedback},'' {\em Optics Letters},
  vol.~12, no.~11, p.~876, 1987.

\bibitem{Notcutt:2012ca}
M.~Notcutt, ``{Frequency Stabilization to Hz-level linewidths using Fabry-Perot
  Cavities},'' {\em Conference on Lasers and Electro-Optics 2012}, 2012.

\bibitem{Salomon:1988fc}
C.~Salomon, D.~Hils, and J.~L. Hall, ``{Laser stabilization at the millihertz
  level},'' {\em Journal of the Optical Society of America B}, vol.~5, no.~8,
  p.~1576, 1988.

\bibitem{III:1976bp}
J.~W. Berthold~III, S.~F. Jacobs, and M.~A. Norton, ``{Dimensional stability of
  fused silica, Invar, and several ultralow thermal expansion materials},''
  {\em Applied Optics}, vol.~15, no.~8, p.~1898, 1976.

\bibitem{Takahashi:2012bh}
A.~Takahashi, ``{Long-term dimensional stability of a line scale made of low
  thermal expansion ceramic NEXCERA},'' {\em Measurement Science and
  Technology}, vol.~23, no.~3, p.~035001, 2012.

\bibitem{Dawkins:2008ef}
S.~T. Dawkins and A.~N. Luiten, ``{Single actuator alignment control for
  improved frequency stability of a cavity-based optical frequency
  reference},'' {\em Applied Optics}, vol.~47, no.~9, p.~1239, 2008.

\bibitem{Keller:2013dm}
J.~Keller, S.~Ignatovich, S.~A. Webster, and T.~E. Mehlst{\"a}ubler, ``{Simple
  vibration-insensitive cavity for laser stabilization at the $10^{-16}$
  level},'' {\em Applied Physics B}, vol.~116, no.~1, p.~203, 2013.

\bibitem{Braginsky:2003cp}
V.~B. Braginsky and S.~P. Vyatchanin, ``{Thermodynamical fluctuations in
  optical mirror coatings},'' {\em Physics Letters A}, vol.~312, no.~3-4,
  p.~244, 2003.

\bibitem{Thorpe:2008bv}
J.~I. Thorpe, K.~Numata, and J.~Livas, ``{Laser frequency stabilization and
  control through offset sideband locking to optical cavities},'' {\em Optics
  Express}, vol.~16, no.~20, p.~15980, 2008.

\bibitem{Siegman:1986uk}
A.~E. Siegman, {\em {Lasers}}.
\newblock University Science Books, 1986.

\bibitem{Drever:1983do}
R.~W.~P. Drever, J.~L. Hall, F.~V. Kowalski, J.~Hough, G.~M. Ford, A.~J.
  Munley, and H.~Ward, ``{Laser phase and frequency stabilization using an
  optical resonator},'' {\em Applied Physics B Photophysics and Laser
  Chemistry}, vol.~31, no.~2, p.~97, 1983.

\bibitem{Black:2001fe}
E.~D. Black, ``{An introduction to Pound{\textendash}Drever{\textendash}Hall
  laser frequency stabilization},'' {\em American Journal of Physics}, vol.~69,
  no.~1, p.~79, 2001.

\bibitem{Houssin:1990jr}
M.~Houssin, M.~Jardino, and M.~Desaintfuscien, ``{Comparison of the calculated
  transient responses of a Fabry{\textendash}Perot used in reflection and in
  transmission},'' {\em Review of Scientific Instruments}, vol.~61, no.~11,
  p.~3348, 1990.

\bibitem{Kobayashi:1982ij}
S.~Kobayashi, Y.~Yamamoto, M.~Ito, and T.~Kimura, ``{Direct frequency
  modulation in AlGaAs semiconductor lasers},'' {\em IEEE Journal of Quantum
  Electronics}, vol.~18, no.~4, p.~582, 1982.

\bibitem{Saito:1984gp}
S.~Saito, O.~Nilsson, and Y.~Yamamoto, ``{Coherent FSK transmitter using a
  negative feedback stabilised semiconductor laser},'' {\em Electronics
  Letters}, vol.~20, no.~17, p.~703, 1984.

\bibitem{Corrc:1994bc}
P.~Corrc, O.~Girad, and I.~F. de~Faria, ``{On the thermal contribution to the
  FM response of DFB lasers: theory and experiment},'' {\em IEEE Journal of
  Quantum Electronics}, vol.~30, no.~11, p.~2485, 1994.

\bibitem{Takamoto:2005ef}
M.~Takamoto, F.-L. Hong, R.~Higashi, and H.~Katori, ``{An optical lattice
  clock},'' {\em Nature}, vol.~435, no.~7040, p.~321, 2005.

\bibitem{Fleischhauer:2005ix}
M.~Fleischhauer, A.~Imamoglu, and J.~P. Marangos, ``{Electromagnetically
  induced transparency: Optics in coherent media},'' {\em Reviews of Modern
  Physics}, vol.~77, no.~2, p.~633, 2005.

\bibitem{Mack:2011he}
M.~Mack, F.~Karlewski, H.~Hattermann, S.~H{\"o}ckh, F.~Jessen, D.~Cano, and
  J.~Fort{\'a}gh, ``{ Measurement of absolute transition frequencies of Rb87 to
  nS and nD Rydberg states by means of electromagnetically induced transparency
  },'' {\em Physical Review A}, vol.~83, no.~5, 2011.

\bibitem{Tauschinsky:2013iq}
A.~Tauschinsky, R.~Newell, H.~B. v.~L. van~den Heuvell, and R.~J.~C. Spreeuw,
  ``{ Measurement of 87 Rb Rydberg-state hyperfine splitting in a
  room-temperature vapor cell },'' {\em Physical Review A}, vol.~87, no.~4,
  2013.

\bibitem{Gea-Banacloche:1995ix}
J.~Gea-Banacloche, Y.-q. Li, S.-z. Jin, and M.~Xiao, ``{Electromagnetically
  induced transparency in ladder-type inhomogeneously broadened media: Theory
  and experiment},'' {\em Physical Review A}, vol.~51, no.~1, p.~576, 1995.

\bibitem{Elliott:1982jm}
D.~S. Elliott, R.~Roy, and S.~J. Smith, ``{Extracavity laser band-shape and
  bandwidth modification},'' {\em Physical Review A}, vol.~26, no.~1, p.~12,
  1982.

\bibitem{Saito:1981fz}
S.~Saito and Y.~Yamamoto, ``{Direct observation of Lorentzian lineshape of
  semiconductor laser and linewidth reduction with external grating
  feedback},'' {\em Electronics Letters}, vol.~17, no.~9, p.~325, 1981.

\bibitem{Barnes:1971ik}
J.~A. Barnes, A.~R. Chi, L.~S. Cutler, D.~J. Healey, D.~B. Leeson, T.~E.
  McGunigal, J.~A. Mullen, W.~L. Smith, R.~L. Sydnor, R.~F.~C. Vessot, and
  G.~M.~R. Winkler, ``{Characterization of Frequency Stability},'' {\em IEEE
  Transactions on Instrumentation and Measurement}, vol.~IM-20, no.~2, p.~105,
  1971.

\bibitem{Okoshi:1980kt}
T.~Okoshi, K.~Kikuchi, and A.~Nakayama, ``{Novel method for high resolution
  measurement of laser output spectrum},'' {\em Electronics Letters}, vol.~16,
  no.~16, p.~630, 1980.

\bibitem{Horak:2006ja}
P.~Horak and W.~H. Loh, ``{On the delayed self-heterodyne interferometric
  technique for determining the linewidth of fiber lasers},'' {\em Optics
  Express}, vol.~14, no.~9, p.~3923, 2006.

\bibitem{Domenico:2010icbaca}
G.~Di~Domenico, S.~Schilt, and P.~Thomann, ``{Simple approach to the relation
  between laser frequency noise and laser line shape},'' {\em Applied Optics},
  vol.~49, no.~25, p.~4801, 2010.

\bibitem{Ishida:1991kk}
O.~Ishida, ``{Delayed-self-heterodyne measurement of laser frequency
  fluctuations},'' {\em Journal of Lightwave Technology}, vol.~9, no.~11,
  p.~1528, 1991.

\bibitem{Dumke:2002hn}
R.~Dumke, M.~Volk, T.~M{\"u}ther, F.~B.~J. Buchkremer, G.~Birkl, and W.~Ertmer,
  ``{Micro-optical Realization of Arrays of Selectively Addressable Dipole
  Traps: A Scalable Configuration for Quantum Computation with Atomic
  Qubits},'' {\em Physical Review Letters}, vol.~89, no.~9, 2002.

\bibitem{Zoubi:2014hz}
H.~Zoubi, ``{Collective interactions in an array of atoms coupled to a
  nanophotonic waveguide},'' {\em Physical Review A}, vol.~89, no.~4, 2014.

\bibitem{Akbulut:2011fn}
D.~Akbulut, T.~J. Huisman, E.~G. van Putten, W.~L. Vos, and A.~P. Mosk,
  ``{Focusing light through random photonic media by binary amplitude
  modulation},'' {\em Optics Express}, vol.~19, no.~5, p.~4017, 2011.

\bibitem{Muldoon:2012jt}
C.~Muldoon, L.~Brandt, J.~Dong, D.~Stuart, E.~Brainis, M.~Himsworth, and
  A.~Kuhn, ``{Control and manipulation of cold atoms in optical tweezers},''
  {\em New Journal of Physics}, vol.~14, no.~7, p.~073051, 2012.

\bibitem{Lee:1974ed}
W.-H. Lee, ``{Binary Synthetic Holograms},'' {\em Applied Optics}, vol.~13,
  no.~7, p.~1677, 1974.

\bibitem{Conkey:2012cr}
D.~B. Conkey, A.~M. Caravaca-Aguirre, and R.~Piestun, ``{High-speed scattering
  medium characterization with application to focusing light through turbid
  media},'' {\em Optics Express}, vol.~20, no.~2, p.~1733, 2012.

\bibitem{Bakr:2009bx}
W.~S. Bakr, J.~I. Gillen, A.~Peng, S.~F{\"o}lling, and M.~Greiner, ``{A quantum
  gas microscope for detecting single atoms in a Hubbard-regime optical
  lattice},'' {\em Nature}, vol.~462, no.~7269, p.~74, 2009.

\bibitem{Bijnen:2015da}
R.~M.~W. van Bijnen, C.~Ravensbergen, D.~J. Bakker, G.~J. Dijk, S.~J. J. M.~F.
  Kokkelmans, and E.~J.~D. Vredenbregt, ``{Patterned Rydberg excitation and
  ionization with a spatial light modulator},'' {\em New Journal of Physics},
  vol.~17, no.~2, p.~023045, 2015.

\bibitem{Bergamini:2004bl}
S.~Bergamini, B.~Darqui{\'e}, M.~Jones, L.~Jacubowiez, A.~Browaeys, and
  P.~Grangier, ``{Holographic generation of micro-trap arrays for single
  atoms},'' {\em Conference on Lasers and Electro-Optics/International Quantum
  Electronics Conference and Photonic Applications Systems Technologies}, 2004.

\bibitem{Lazarev:2012ku}
G.~Lazarev, A.~Hermerschmidt, S.~Kr{\"u}ger, and S.~Osten, {\em {Optical
  Imaging and Metrology: Advanced Technologies}}.
\newblock Wiley-VCH Verlag GmbH {\&} Co, 2012.

\bibitem{Gunther:2004ix}
W.~Gunther, K.~Sven, K.~Jorn, G.~Hartmut, D.~Nazif, D.~Matthias, and
  T.~Stephan, ``{Application of a Liquid Crystal Display Spatial Light
  Modulator System as Dynamic Diffractive Element and in Optical Image
  Processing},'' {\em Journal of Optical Communications}, vol.~25, no.~4, 2004.

\bibitem{Ramanath:2015ua}
R.~Ramanath, ``{Digital Micromirror Device and Digital Light Processing},''
  {\em Handbook of Digital Imaging}, pp.~1--18, 2015.

\bibitem{Knipe:1996gz}
R.~L. Knipe, ``{Challenges of a Digital Micromirror Device: modeling and
  design},'' {\em Micro-Optical Technologies for Measurement, Sensors, and
  Microsystems}, 1996.

\bibitem{Nogrette:2014fj}
F.~Nogrette, H.~Labuhn, S.~Ravets, D.~Barredo, L.~B{\'e}guin, A.~Vernier,
  T.~Lahaye, and A.~Browaeys, ``{Single-Atom Trapping in Holographic 2D Arrays
  of Microtraps with Arbitrary Geometries},'' {\em Physical Review X}, vol.~4,
  no.~2, 2014.

\bibitem{Serati:2005dg}
S.~Serati and J.~Stockley, ``{Advances in liquid crystal based devices for
  wavefront control and beamsteering},'' {\em Advanced Wavefront Control:
  Methods, Devices, and Applications III}, 2005.

\bibitem{Bryngdahl:1975jv}
O.~Bryngdahl, ``{Optical scanner - light deflection using computer-generated
  diffractive elements},'' {\em Optics Communications}, vol.~15, p.~237, Jan.
  1975.

\bibitem{Wyrowski:1988dh}
F.~Wyrowski and O.~Bryngdahl, ``{Iterative Fourier-transform algorithm applied
  to computer holography},'' {\em Journal of the Optical Society of America A},
  vol.~5, no.~7, p.~1058, 1988.

\bibitem{Bijnen:2013ve}
K.~A. van Leeuwen and R.~M.~W. van Bijnen, {\em {Quantum Engineering with
  Ultracold Atoms}}.
\newblock PhD thesis, Technische Universiteit Eindhoven, 2013.

\bibitem{Huignard:1987go}
J.~P. Huignard, ``{Spatial light modulators and their applications},'' {\em
  Journal of Optics}, vol.~18, no.~4, p.~181, 1987.

\bibitem{Banyal:2010fz}
R.~K. Banyal and B.~R. Prasad, ``{Nonlinear response studies and corrections
  for a liquid crystal spatial light modulator},'' {\em Pramana}, vol.~74,
  no.~6, p.~961, 2010.

\bibitem{Neff:1990fg}
J.~A. Neff, R.~A. Athale, and S.~H. Lee, ``{Two-dimensional spatial light
  modulators: a tutorial},'' {\em Proceedings of the IEEE}, vol.~78, no.~5,
  p.~826, 1990.

\bibitem{Leung:2014gw}
V.~Y.~F. Leung, D.~R.~M. Pijn, H.~Schlatter, L.~Torralbo-Campo, A.~L. La~Rooij,
  G.~B. Mulder, J.~Naber, M.~L. Soudijn, A.~Tauschinsky, C.~Abarbanel,
  B.~Hadad, E.~Golan, R.~Folman, and R.~J.~C. Spreeuw, ``{Magnetic-film atom
  chip with 10 $\mu$m period lattices of microtraps for quantum information
  science with Rydberg atoms},'' {\em Review of Scientific Instruments},
  vol.~85, no.~5, p.~053102, 2014.

\bibitem{Lizana:2009ia}
A.~Lizana, N.~Mart{\'\i}n, M.~Estap{\'e}, E.~Fern{\'a}ndez, I.~Moreno,
  A.~M{\'a}rquez, C.~Iemmi, J.~Campos, and M.~J. Yzuel, ``{Influence of the
  incident angle in the performance of Liquid Crystal on Silicon displays},''
  {\em Optics Express}, vol.~17, no.~10, p.~8491, 2009.

\bibitem{Jesacher:2007cv}
A.~Jesacher, A.~Schwaighofer, S.~F{\"u}rhapter, C.~Maurer, S.~Bernet, and
  M.~Ritsch-Marte, ``{Wavefront correction of spatial light modulators using an
  optical vortex image},'' {\em Optics Express}, vol.~15, no.~9, p.~5801, 2007.

\bibitem{García-Márquez:2011jh}
J.~Garc{\'\i}a-M{\'a}rquez, J.~E. Landgrave, N.~Alcal{\'a}-Ochoa, and
  C.~P{\'e}rez-Santos, ``{Recursive wavefront aberration correction method for
  LCoS spatial light modulators},'' {\em Optics and Lasers in Engineering},
  vol.~49, no.~6, p.~743, 2011.

\bibitem{Hu:2007cf}
Q.~Hu and K.~G. Harding, ``{Conversion from phase map to coordinate: Comparison
  among spatial carrier, Fourier transform, and phase shifting methods},'' {\em
  Optics and Lasers in Engineering}, vol.~45, no.~2, p.~342, 2007.

\bibitem{Carré:1966gy}
P.~Carr{\'e}, ``{Installation et utilisation du comparateur photo{\'e}lectrique
  et interf{\'e}rentiel du Bureau International des Poids et Mesures},'' {\em
  Metrologia}, vol.~2, no.~1, p.~13, 1966.

\bibitem{Magalhaes:2010jk}
P.~A.~A. Magalhaes, P.~S. Neto, and C.~A. Magalh{\~a}es, ``{A modified carre
  algorithm for phase shifting interferometry},'' {\em Journal of Optics},
  vol.~39, no.~1, p.~5, 2010.

\bibitem{Amézquita:2011bd}
R.~Am{\'e}zquita, O.~Rinc{\'o}n, and Y.~M. Torres, ``{Aberration compensation
  using a spatial light modulator LCD},'' {\em Journal of Physics: Conference
  Series}, vol.~274, p.~012111, 2011.

\bibitem{Malacara:2007tk}
D.~Malacara, {\em {Optical Shop Testing}}.
\newblock Wiley-Interscience, 2007.

\bibitem{Huntley:1989kt}
J.~M. Huntley, ``{Noise-immune phase unwrapping algorithm},'' {\em Applied
  Optics}, vol.~28, no.~16, p.~3268, 1989.

\bibitem{Buckland:1995ko}
J.~R. Buckland, J.~M. Huntley, and S.~R.~E. Turner, ``{Unwrapping noisy phase
  maps by use of a minimum-cost-matching algorithm},'' {\em Applied Optics},
  vol.~34, no.~23, p.~5100, 1995.

\bibitem{Hardy:2000kk}
J.~W. Hardy and L.~Thompson, ``{Adaptive Optics for Astronomical Telescopes},''
  {\em Physics Today}, vol.~53, no.~4, p.~69, 2000.

\bibitem{Kubby:2013fu}
J.~Kubby, X.~Tao, and O.~Azucena, ``{Adaptive Optics for Biological Imaging
  using Direct Wavefront Sensing},'' {\em Imaging and Applied Optics}, 2013.

\bibitem{Shack:1971vy}
R.~V. Shack and B.~C. Platt, ``{Program of the 1971 Spring Meeting of the
  Optical Society of America},'' {\em JOSA}, vol.~61, pp.~656--658, May 1971.

\bibitem{PEARSON:1979kf}
J.~E. Pearson, R.~H. Freeman, and H.~C. Reynolds, ``{Adaptive Optical
  Techniques for Wave-Front Correction},'' {\em Applied Optics and Optical
  Engineering}, p.~245, 1979.

\bibitem{Fugate:1994kx}
R.~Q. Fugate, B.~L. Ellerbroek, C.~H. Higgins, M.~P. Jelonek, W.~J. Lange,
  A.~C. Slavin, W.~J. Wild, D.~M. Winker, , J.~M. Wynia, J.~M. Spinhirne, B.~R.
  Boeke, R.~E. Ruane, J.~F. Moroney, , M.~D. Oliker, D.~W. Swindle, and R.~A.
  Cleis, ``{Two generations of laser-guide-star adaptive-optics experiments at
  the Starfire Optical Range},'' {\em Journal of the Optical Society of America
  A}, vol.~11, no.~1, p.~310, 1994.

\bibitem{Liang:1994dp}
J.~Liang, B.~Grimm, S.~Goelz, and J.~F. Bille, ``{Objective measurement of wave
  aberrations of the human eye with the use of a Hartmann-Shack wave-front
  sensor},'' {\em Journal of the Optical Society of America A}, vol.~11, no.~7,
  p.~1949, 1994.

\bibitem{Bowman:2010dq}
R.~W. Bowman, A.~J. Wright, and M.~J. Padgett, ``{An SLM-based
  Shack{\textendash}Hartmann wavefront sensor for aberration correction in
  optical tweezers},'' {\em Journal of Optics}, vol.~12, no.~12, p.~124004,
  2010.

\bibitem{Southwell:1980fc}
W.~H. Southwell, ``{Wave-front estimation from wave-front slope
  measurements},'' {\em Journal of the Optical Society of America}, vol.~70,
  no.~8, p.~998, 1980.

\bibitem{Topa:2002dt}
D.~M. Topa, ``{Wavefront reconstruction for the Shack-Hartmann wavefront
  sensor},'' {\em Optical Design and Analysis Software II}, 2002.

\bibitem{Mahajan:2007ko}
V.~N. Mahajan and G.-m. Dai, ``{Orthonormal polynomials in wavefront analysis:
  analytical solution},'' {\em Journal of the Optical Society of America A},
  vol.~24, no.~9, p.~2994, 2007.

\end{thebibliography}
